\documentclass[conference]{IEEEtran}
\usepackage[numbers,sort&compress]{natbib}
\usepackage[utf8]{inputenc}
\usepackage[T1]{fontenc}
\PassOptionsToPackage{hyphens}{url}
\usepackage{hyperref}
\hypersetup{hidelinks}
\usepackage{url}
\usepackage{xurl}
\DeclareFontShape{T1}{ptm}{m}{scit}{<-> ssub * ptm/m/sc}{}
\usepackage{booktabs}
\usepackage{amsfonts}
\usepackage{amsmath}
\usepackage{amssymb}
\usepackage{amsthm}
\usepackage{nicefrac}
\usepackage{microtype}
\usepackage{iftex}
\usepackage{xcolor}
\usepackage{graphicx}
\usepackage{enumitem}
\usepackage{array}
\usepackage{placeins}
\usepackage{tikz}
\usetikzlibrary{positioning, arrows.meta, calc, fit, backgrounds, shapes.geometric, shapes.multipart}
\newcommand{\sudp}{\textsc{SUDP}}
\newcommand{\eg}{\textit{e.g.}}

\newtheoremstyle{paper}%
  {4pt}{4pt}{\itshape}{}{\bfseries}{.}{.5em}%
  {\thmname{#1}~\thmnumber{#2}\thmnote{ (#3)}}
\theoremstyle{paper}
\newtheorem{definition}{Definition}
\newtheorem{theorem}{Theorem}
\newtheorem{proposition}{Proposition}
\newtheorem{lemma}{Lemma}

\title{\sudp: Secret-Use Delegation Protocol\\for Agentic Systems}
\IEEEoverridecommandlockouts
\author{%
\IEEEauthorblockN{Xiaohang Yu}
\IEEEauthorblockA{Imperial College London\\
x.yu21@imperial.ac.uk}
\and
\IEEEauthorblockN{Hejia Geng}
\IEEEauthorblockA{University of Oxford\\
genghejia0530@gmail.com}
\and
\IEEEauthorblockN{Xinmeng Zeng}
\IEEEauthorblockA{Stanford University\\
xinzeng@stanford.edu}
\and
\IEEEauthorblockN{William Knottenbelt\IEEEauthorrefmark{1}}
\IEEEauthorblockA{Imperial College London\\
w.knottenbelt@imperial.ac.uk}
\thanks{\IEEEauthorrefmark{1}Corresponding author.}
}

\newcommand{\artifactlink}{\url{https://github.com/xhyumiracle/sudp}}
\newcommand{\artifactnoun}{reference implementation}

\begin{document}

\maketitle

\begin{abstract}
Agentic systems now act with user secrets for APIs, messaging platforms, and cloud services. Today's agent runtimes typically implement \emph{authorization by exposure}: enabling action means placing a reusable secret, or a reusable artifact derived from it, inside the runtime, so a transient prompt-injection or tool-side compromise becomes durable account compromise. Existing defenses cover adjacent pieces such as secret storage, scoped delegation, sender-constrained tokens, and runtime monitoring, but none states the combined requirement: an untrusted autonomous requester should be able to cause a user-authorized secret-backed operation without gaining reusable authority over it. We formalize this as the \emph{Agent Secret Use} (ASU) problem and identify seven security properties any solution must satisfy, spanning authorization integrity and secret confidentiality. We propose the \emph{Secret-Use Delegation Protocol} (\sudp{}), in which a requester proposes a canonical operation, the user authorizes it with a fresh authenticator-backed grant, and a custodian redeems the grant to perform the bounded use; reusable authority never crosses the requester boundary. We specialize \sudp{} for LLM-driven agents, where it applies whenever a tool call would exercise user-enrolled authority-bearing material. Under standard cryptographic assumptions, \sudp{} satisfies six of the seven properties; the seventh, runtime-memory confidentiality, is closed by composition with a hardware-rooted runtime. A reference implementation, available at \artifactlink{}, adds sub-millisecond protocol overhead per mediated call.
\end{abstract}

\section{Introduction}
\label{sec:intro}

A coding agent holding a GitHub token can be steered by an injected README into pushing code to a public repository: once a reusable secret is inside the agent runtime, indirect prompt injection becomes durable account compromise~\citep{greshake2023indirect,owasp2024llm}. Tool-augmented language-model agents now routinely perform such authority-bearing actions: they call model and platform APIs~\citep{toolformer2023,mialon2023augmented,yao2023react}, send messages, and authenticate to third-party tools~\citep{langchain2024,openai2025agentssdk}, backed by API keys, OAuth tokens, or signing keys held in environment variables, databases, or secret managers~\citep{hashicorp2024vault,aws2024secretsmanager} and handed to the agent for ad-hoc use. Across successive agent-safety benchmarks~\citep{zhan2024injecagent,zhang2024agentsafetybench,debenedetti2024agentdojo,zhou2025haicosystem}, agent runtimes have continued to fall to prompt injection and tool misuse, including on secret-adjacent tasks, despite improving base models; under the bearer-secret pattern, every such failure becomes durable authority compromise.

The question for agentic systems is: \emph{how can an agent be authorized to cause a secret-backed operation without ever gaining reusable authority over it?} Today's model bundles capability and possession into a single artifact, a pattern we call \emph{authorization by exposure}. Existing approaches address neighboring dimensions: capability-based authority~\citep{dennis1966programming,miller2006robust}, scoped delegation~\citep{hardt2012oauth,birgisson2014macaroons}, sender-constrained bearers~\citep{fett2023dpop}, user-mediated authenticators with derivable material~\citep{webauthn2019,fido2prf2023}, and standardized cryptographic components~\citep{krawczyk2010hmac,dworkin2012keywrap,schaad2002keywrap,housley2009kwp,barnes2022hpke,rescorla2018tls13}; none provides a common specification for \emph{authorization without giving the requester reusable authority}. We compose these elements into a unified protocol whose unit of delegation is one authorized operation rather than the secret itself. The safety of this authorization layer, as much as reasoning ability, bounds what agents can be allowed to do autonomously.

\begin{figure}[t]
\centering
\includegraphics[width=0.88\columnwidth]{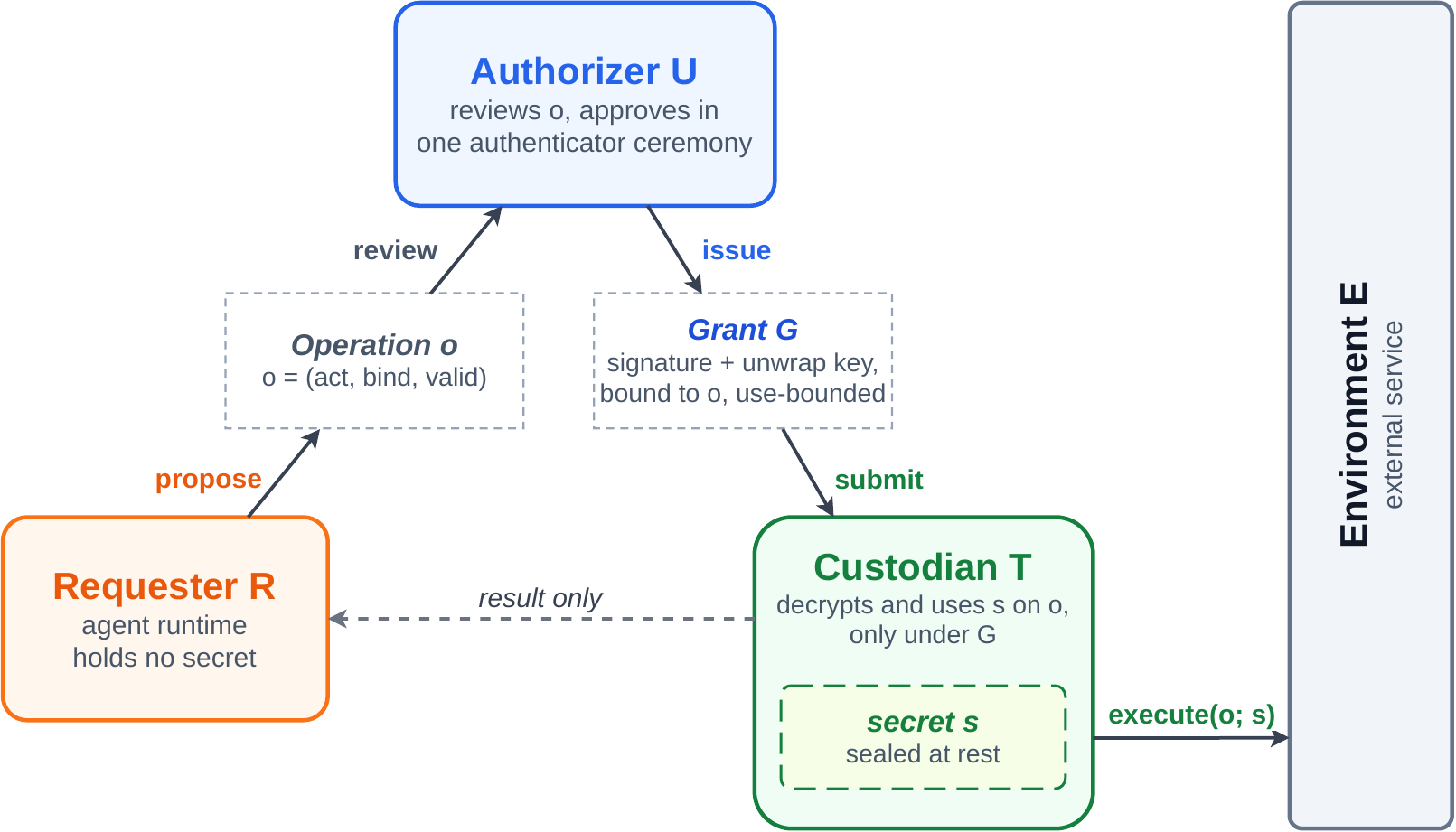}
\caption{Schematic of \sudp{}. The requester $R$ proposes the operation $o$; the authorizer $U$ reviews $o$ and issues the grant $G$, signature-bound to $\mathsf{H}(o)$ and freshness $r$, carrying the unwrap key for the secret $s$; the custodian $T$ redeems $G$ and spends $s$ at the environment $E$. Only the result returns to $R$: reusable authority never crosses the requester boundary (\S\ref{sec:protocol}).}
\label{fig:arch}
\end{figure}

\paragraph{Contributions.}
\textbf{(i)~The Agent Secret Use problem} (\S\ref{sec:agentsecretuse}). We formalize Agent Secret Use and define seven security properties that any solution must satisfy. The taxonomy admits uniform evaluation of secret-mediation, delegation-token, and TEE-backed systems, which \S\ref{sec:positioning} carries out.

\textbf{(ii)~The \sudp{} protocol} (\S\ref{sec:protocol}). We propose the Secret-Use Delegation Protocol as a solution to ASU, with one authorized operation rather than the secret as the unit of delegation, and prove, under standard cryptographic assumptions, that it satisfies six of the seven properties natively; the seventh is closed by composition with a hardware-rooted runtime (\S\ref{sec:security}).

\textbf{(iii)~Agentic application and instantiation} (\S\ref{sec:interface}--\ref{sec:concrete}). We describe how \sudp{} applies in an LLM-based agent system, where even a fully prompt-injected agent can at most propose operations subject to explicit authorization, and instantiate the protocol with standard cryptographic components, introducing no new assumption. A reference implementation of the protocol core accompanies the paper.

\section{Related Work}
\label{sec:related}

\paragraph{Agent systems and external capabilities.}
AutoGPT and AutoGen made long-horizon, tool-using agents a concrete software pattern rather than only a prompting style~\citep{autogpt2023,autogen2023}. Recent products push the same pattern toward browser, desktop, and multi-step task execution: Manus, OpenAI Operator, and Anthropic's computer-use interface~\citep{manus2025,openai2025operator,anthropic2024computeruse}.

\paragraph{Secret management, authorization, and delegation.}
Capability-based access control~\citep{dennis1966programming,miller2006robust}, OAuth~2.0 bearers~\citep{hardt2012oauth,rfc6750}, DPoP~\citep{fett2023dpop}, OAuth Rich Authorization Requests~\citep{ietf2025rar}, GNAP~\citep{ietf2024gnap}, MCP authorization~\citep{mcp2025auth}, and Macaroons~\citep{birgisson2014macaroons} address scoped delegation but pass possession-bearing or reusable artifacts through the requester. WebAuthn~\citep{webauthn2019} and its PRF extension~\citep{fido2prf2023} provide phishing-resistant user authentication and authenticator-bound key derivation, primitives that \sudp{} builds on (\S\ref{sec:protocol}). Secret managers and KMS services~\citep{hashicorp2024vault,aws2024secretsmanager} provide mature storage, access control, and rotation, but their privileged backend assumes the authorization decision is upstream of the requester; \sudp{} complements them at the authorization layer. A recent line of agentic delegation work extends web-identity and provenance constructs to agent principals: Authenticated Delegation~\citep{south2025authenticateddelegation} binds agent credentials to a verified human delegator; SAGA~\citep{syros2025saga} gates inter-agent contact through a user-registered provider that distributes one-time access-control keys; Agentic JWT~\citep{goswami2025agenticjwt} binds each agent action to verifiable user intent and, optionally, to a workflow step; HDP~\citep{dalugoda2026hdp} targets multi-hop human-delegation provenance; and OIDC-A~\citep{oidca2025} extends OIDC with agent identity. Each touches a different subset of the ASU axes; the per-axis comparison is in \S\ref{sec:positioning-axes}.

\paragraph{Runtime defenses, monitors, and benchmarks.}
A large body of work hardens the agent or contains the consequences of misbehavior: instruction-data separation and prompt-engineering defenses~\citep{chen2024struq,chen2024secalign,hines2024spotlighting}, capability-tracking interpreters~\citep{debenedetti2025camel}, per-tool policy engines~\citep{shi2025progent,wang2026agentspec,tsai2025conseca}, verify-before-commit mitigations~\citep{lin2026vigil}, guardrail stacks~\citep{meta2025llamafirewall,liu2026clawkeeper}, and execution sandboxes~\citep{wu2025secgpt}. These approaches reason about what the agent does or contain where it does it; \sudp{} constrains what a secret can be used for and composes with them rather than substituting. Recent adaptive evaluations show that stronger attacks bypass published behavioral defenses~\citep{nasr2025attackermovessecond}, motivating protocol-level guarantees beyond behavioral filtering, and an empirical audit of 17,022 deployed LLM agent skills finds exploitable and persistent credential leakage in 520 of them~\citep{chen2026credleakage}. Agent-safety benchmarks across model generations report nontrivial failure rates: attack success under prompt injection~\citep{zhan2024injecagent,debenedetti2024agentdojo,zhang2024asb}, unsafe tool use~\citep{ruan2023toolemu,zhou2025haicosystem,vijayvargiya2025openagentsafety}, and privacy leakage across tool and inter-agent channels~\citep{qiao2025topbench,agentleak2026}. Surveys identify ad-hoc secret handling and least-privilege maintenance over credentials as an open gap~\citep{kim2026attackdefenselandscape,li2026securityconsiderations}; lifecycle analyses place capability enforcement at the execution stage~\citep{deng2026tamingopenclaw,ying2026fasa}.

\section{Agent Secret Use}
\label{sec:agentsecretuse}

\emph{Agent Secret Use} (ASU) arises when an autonomous requester must cause a user-authorized operation without ever gaining reusable authority over that operation.

\subsection{The Problem}
\label{sec:problem}
\label{sec:trust-model}

\begin{figure}[t]
\centering
\includegraphics[width=\columnwidth]{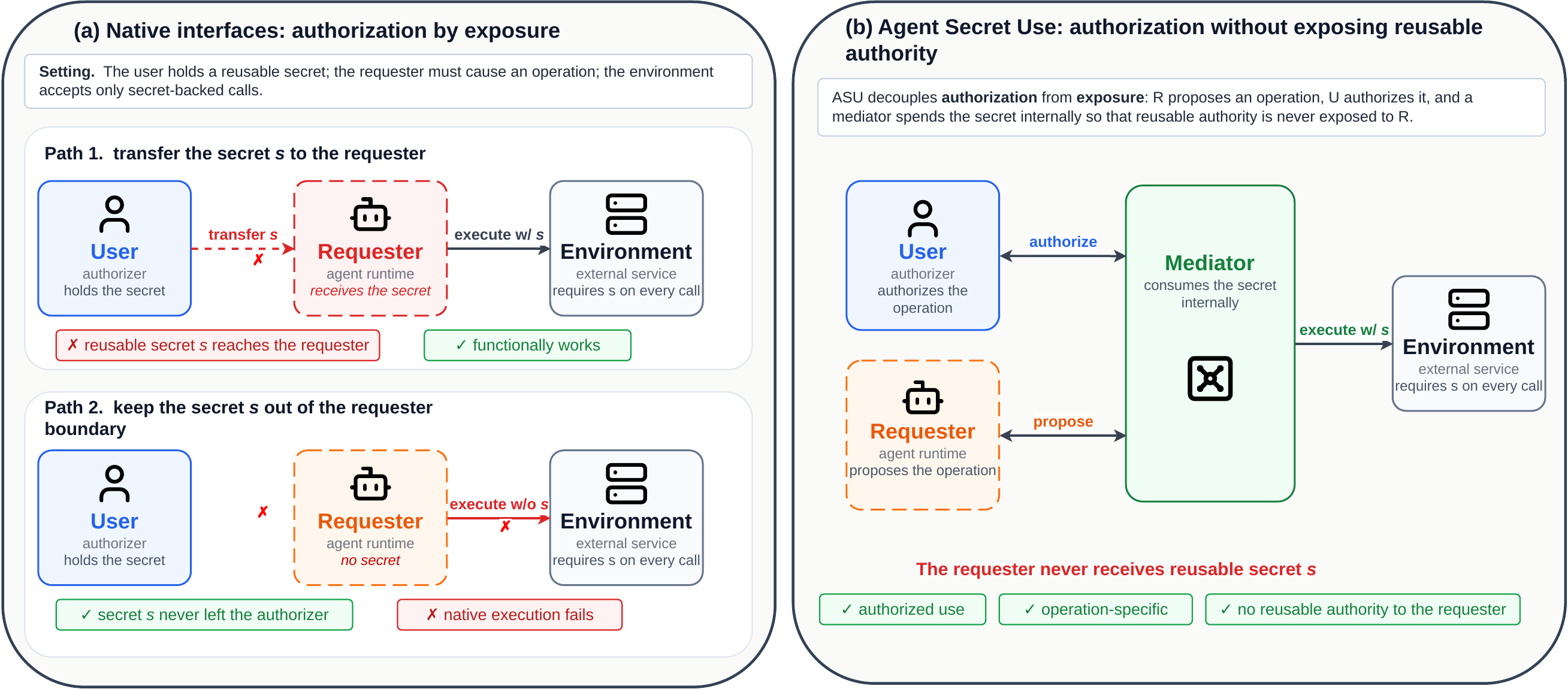}
\caption{\textbf{Agent Secret Use as a structural problem.} \textbf{(a)}~Native secret-backed interfaces implement \emph{authorization by exposure}: handing $s$ to $R$ functionally works but gives $R$ reusable authority over operations at $E$; keeping $s$ outside $R$ preserves confidentiality but the native call is rejected by $E$. \textbf{(b)}~ASU decouples authorization from exposure: $R$ proposes the operation, $U$ authorizes it, and $s$ stays inside a trust boundary outside $R$, so $R$ never gains reusable authority over operations at $E$.}
\label{fig:asu}
\end{figure}

Agent Secret Use poses a conflict between three parties. A user \(U\) holds authority at an external environment \(E\), exercised through authority-bearing material \(s\) (Definition~\ref{def:authority-bearing}) that \(E\) accepts on \(U\)'s behalf for operations in a space \(\mathcal{O}_E\). An autonomous requester \(R\), acting on \(U\)'s behalf, lies outside \(U\)'s trust boundary: \(R\) must never gain reusable authority (Definition~\ref{def:reusable-authority}) over those operations. The unmediated native interface cannot satisfy both requirements (Figure~\ref{fig:asu}); Definition~\ref{def:agentsecretuse} states the resulting problem.

\begin{definition}[Authority-bearing material]\label{def:authority-bearing}
Material \(s\) is \emph{authority-bearing at \(E\)} if possession of \(s\), or of material derived from \(s\) that \(E\) accepts, enables operations at \(E\) under the authority of the party that enrolled \(s\).
\end{definition}

\begin{definition}[Reusable authority]\label{def:reusable-authority}
\emph{Reusable authority} over operations at \(E\) is the capability to cause more than one invocation of operations at \(E\), whether via direct possession of authority-bearing material (Definition~\ref{def:authority-bearing}) or via any derived artifact \(E\) accepts.
\end{definition}

\begin{definition}[Agent Secret Use Problem]\label{def:agentsecretuse}
\emph{Agent Secret Use} is the problem of enabling an untrusted autonomous requester \(R\), acting on \(U\)'s behalf, to cause operations in \(\mathcal{O}_E\) to be executed at \(E\), without ever exposing \(R\) to reusable authority (Definition~\ref{def:reusable-authority}) over what \(U\) explicitly authorizes.
\end{definition}

\subsection{Adversary Model}
\label{sec:adversary}

We parameterize the adversary \(\mathcal{A}\) by capability:
\begin{itemize}[leftmargin=*,topsep=2pt,itemsep=1pt]
  \item \(\mathcal{A}^{R}\): controls \(R\).
  \item \(\mathcal{A}^{\mathrm{net}}\): controls the network (observes, modifies, drops, replays, and reorders any message).
  \item \(\mathcal{A}^{\mathrm{state}}\): reads at rest the persistent state of any component a solution introduces to hold authority-bearing material.
  \item \(\mathcal{A}^{\mathrm{mem}}\): reads the runtime memory of such a component during an in-progress operation.
\end{itemize}
\(\mathcal{A}\) may also query \(U\) on any \(o \in \mathcal{O}_E\); \(U\) returns either a decision attributable to \(U\) or a refusal, per a policy the model leaves abstract.

The ASU problem (Definition~\ref{def:agentsecretuse}) is stated against the baseline \((\mathcal{A}^{R}, \mathcal{A}^{\mathrm{net}})\); \(\mathcal{A}^{\mathrm{state}}\) and \(\mathcal{A}^{\mathrm{mem}}\) are robustness extensions targeted by the corresponding axes of \S\ref{sec:taxonomy}.

\paragraph{Goal.}
\(\mathcal{A}\) wins by either \emph{extraction}, obtaining authority-bearing material (Definition~\ref{def:authority-bearing}) and thereby reusable authority (Definition~\ref{def:reusable-authority}) over operations at \(E\), or \emph{misuse}, causing \(E\) to execute some \(o\) that \(U\) did not explicitly authorize. \S\ref{sec:taxonomy} formalizes the security properties that rule these out.

\subsection{Security Properties}
\label{sec:taxonomy}

We formalize the Agent Secret Use problem (Definition~\ref{def:agentsecretuse}) via seven security properties, grouped along the classical confidentiality / integrity distinction~\citep{mccumber1991,saltzer1975protection}. The four \emph{core} properties (AV, OB, UB, CRC) together preclude the two adversarial wins of \S\ref{sec:adversary}: AV, OB, and UB rule out \emph{misuse}; CRC rules out \emph{extraction}. We say a mechanism \emph{solves the ASU problem} if it enforces these four core properties under the baseline adversary \((\mathcal{A}^R, \mathcal{A}^{\mathrm{net}})\). The three \emph{robustness extensions} (CSB, CMC, RFS) target the stronger adversaries \(\mathcal{A}^{\mathrm{state}}\) and \(\mathcal{A}^{\mathrm{mem}}\); they are not required for the baseline solution, and a mechanism that holds secret material only within \(U\) satisfies them trivially. For the protocol of \S\ref{sec:protocol}, Appendix~\ref{app:proofs} restates these properties as adversarial games and proves them by reduction.

\paragraph{Authorization integrity.}
\begin{itemize}[leftmargin=*,topsep=2pt,itemsep=1pt]
  \item \textbf{AV: Authorization Verifiability.} Any successful execution induced through the mechanism is covered by an authorization event attributable to \(U\) under the mechanism's stated identity and trust assumptions. An authenticator-bound assertion \(U\) produces for a specific operation provides stronger direct evidence than an identity provider's assertion about \(U\).
  \item \textbf{OB: Operation Binding.} An authorization is bound to a specific operation \(o\); the mechanism does not induce execution of any operation semantically distinct from \(o\).
  \item \textbf{UB: Use Boundedness.} Execution induced through the mechanism cannot exceed the bounds \(U\) authorized.
\end{itemize}

\paragraph{Secret confidentiality.}
\begin{itemize}[leftmargin=*,topsep=2pt,itemsep=1pt]
  \item \textbf{CRC: Confidentiality under Requester Compromise.} Compromise of the requester does not expose authority-bearing material (Definition~\ref{def:authority-bearing}).
  \item \textbf{CSB: Confidentiality under Storage Breach.} Compromise of the persistent state of a secret-holding component does not expose the secret. Encryption whose unlock key resides in the same compromised state does not satisfy CSB.
  \item \textbf{CMC: Confidentiality under Runtime-Memory Compromise.} Compromise of the runtime memory of a secret-processing component does not expose the secret. Hardware-rooted TEEs may satisfy CMC within their threat model~\citep{costan2016intel}; software-only schemes relying on OS process isolation generally do not.
  \item \textbf{RFS: Rotation Forward Secrecy.} Compromise of secret material in one rotation epoch does not expose material in any other epoch (forward and post-compromise secrecy~\citep{unger2015sok,cohngordon2016pcs}). Rotating only the wrapping key while reusing the same authority-bearing secret is epoch-key isolation, not authority-level forward secrecy.
\end{itemize}

\noindent Deployment concerns (client or authenticator compromise, logs and telemetry, IPC, crash dumps, supply-chain code injection, side channels, out-of-band copying) lie outside the seven security axes.

\FloatBarrier
\section{Positioning}
\label{sec:positioning}

\subsection{Relation to Prior Paradigms}
\label{sec:priorclasses}

Agent Secret Use is a confused-deputy setting~\citep{hardy1988confused} with an added confidentiality constraint: the deputy must not gain reusable authority it could be induced to misuse. Prior paradigms place the authority-holding boundary at different points; ASU prescribes no location, only the negative condition that \(R\) stay outside any boundary containing authority-bearing material accepted at \(E\).

\paragraph{Prior paradigms, by boundary placement.}
\begin{itemize}[leftmargin=*,topsep=2pt,itemsep=2pt]
  \item \emph{Token-based authorization} (OAuth~2.0 bearer~\citep{hardt2012oauth,rfc6750} and sender-constrained~\citep{fett2023dpop} tokens) places the boundary at the token itself: possession or proof-of-possession material is passed to the requester, which remains trusted with reusable authority.
  \item \emph{Authorization-server delegation} (OAuth Rich Authorization Requests~\citep{ietf2025rar}, GNAP~\citep{ietf2024gnap}, MCP authorization~\citep{mcp2025auth}) places the boundary at an authorization server, but the minted token still lands at the requester.
  \item \emph{Centralized secret storage} (Vault, KMS services~\citep{hashicorp2024vault,aws2024secretsmanager}) places the boundary at the storage system and addresses access control and rotation, but the authorization decision sits upstream of the requester rather than being bound to specific operations.
  \item \emph{Hardware-rooted boundaries} (HSMs, secure enclaves, TEE-backed runtimes~\citep{costan2016intel,sabt2015tee}) place the boundary inside tamper-resistant hardware that confines signing keys or code; this addresses host-memory confidentiality but does not by itself authorize specific agentic operations.
  \item \emph{Transaction authentication ceremonies} (closest analogue), exemplified by PSD2 dynamic linking, banking 2FA, Secure Payment Confirmation~\citep{w3c2024spc}, and WebAuthn~UV~\citep{webauthn2019} ceremonies, place the boundary at a trusted user-facing client cooperating with a relying party. This is the prior paradigm closest to ASU's operation-bound user approval, but it assumes the user interacts directly with a trusted client rather than through an autonomous agentic requester.
\end{itemize}
A separate body of concurrent work targets the same agentic setting as \sudp{}: \emph{agentic delegation protocols} (Authenticated Delegation~\citep{south2025authenticateddelegation}, SAGA~\citep{syros2025saga}, OIDC-A~\citep{oidca2025}, HDP~\citep{dalugoda2026hdp}, Agentic JWT~\citep{goswami2025agenticjwt}). These are not prior paradigms but contemporary peers; we compare them per-axis in \S\ref{sec:positioning-axes}.

\subsection{Comparison Along the Security Axes}
\label{sec:positioning-axes}

\paragraph{Framework Independence (FI).}
Alongside the seven security axes we add one deployability axis: a defense is \emph{FI-positive} if any LLM-agent runtime can adopt its specification without a specific agent SDK, planner, tool scheduler, or TEE-backed execution architecture. FI is an adoption property, not a security outcome.

\paragraph{What the table evaluates.}
Table~\ref{tab:compare} restricts attention to systems that make secret-confidentiality or authorization claims. Behavioral monitors and safety evaluators~\citep{meta2025llamafirewall,agentauditor2025}, policy DSLs~\citep{shi2025progent,wang2026agentspec,tsai2025conseca}, taint-tracking interpreters~\citep{debenedetti2025camel}, and execution sandboxes~\citep{wu2025secgpt} operate at orthogonal layers (prompt / tool-call / data-flow / cross-app isolation) and are not captured by our axes; they compose with \sudp{} rather than compete.

\begin{table}[t]
\centering
\footnotesize
\caption{Positioning along the seven security axes of \S\ref{sec:taxonomy} plus FI. Symbol semantics and row notes follow the table. Ratings reflect strict ASU definitions against public documentation at the time of writing; they are not a general security ranking of these systems.}
\label{tab:compare}
\setlength{\tabcolsep}{2.5pt}
\begin{tabular}{@{}>{\raggedright\arraybackslash}p{0.36\columnwidth} cccc ccc c@{}}
\toprule
 & \multicolumn{4}{c}{\textbf{Secret conf.}} & \multicolumn{3}{c}{\textbf{Authz.\ integ.}} & \textbf{Dep.} \\
\cmidrule(lr){2-5}\cmidrule(lr){6-8}\cmidrule(lr){9-9}
 & \textbf{CRC} & \textbf{CSB} & \textbf{CMC} & \textbf{RFS} & \textbf{AV} & \textbf{OB} & \textbf{UB} & \textbf{FI} \\
\midrule
Bearer-token baseline & $\times$ & $\times$ & $\times$ & $\times$ & $\times$ & $\times$ & $\times$ & $\checkmark$ \\
\addlinespace
HashiCorp Vault AI-agent identity pattern~\citep{hashicorp2024vault_agents} & $\times$ & $\triangle$ & $\times$ & $\times$ & $\triangle$ & $\times$ & $\times$ & $\checkmark$ \\
Aegis~\citep{aegis2026} & $\checkmark$ & $\triangle$ & $\times$ & $\times$ & $\times$ & $\times$ & $\times$ & $\checkmark$ \\
Claude Agent SDK proxy~\citep{anthropic2026securedeployment} & $\triangle$ & $\times$ & $\times$ & $\times$ & $\times$ & $\times$ & $\times$ & $\checkmark$ \\
IronClaw~\citep{ironclaw2026} & $\checkmark$ & $\checkmark$ & $\triangle$ & $\times$ & $\times$ & $\times$ & $\times$ & $\times$ \\
\addlinespace
Authenticated Delegation~\citep{south2025authenticateddelegation} & $\times$ & $\times$ & $\times$ & $\times$ & $\triangle$ & $\times$ & $\times$ & $\checkmark$ \\
SAGA~\citep{syros2025saga} & $\times$ & $\times$ & $\times$ & $\times$ & $\checkmark$ & $\triangle$ & $\times$ & $\triangle$ \\
OIDC-A~\citep{oidca2025} & $\times$ & $\times$ & $\times$ & $\times$ & $\triangle$ & $\times$ & $\times$ & $\checkmark$ \\
HDP~\citep{dalugoda2026hdp} & $\times$ & $\times$ & $\times$ & $\times$ & $\checkmark$ & $\times$ & $\triangle$ & $\checkmark$ \\
Agentic JWT~\citep{goswami2025agenticjwt} & $\times$ & $\times$ & $\times$ & $\times$ & $\triangle$ & $\triangle$ & $\triangle$ & $\checkmark$ \\
\midrule
\textbf{\sudp{} (this paper)} & $\checkmark$ & $\checkmark$ & $\triangle$ & $\checkmark$ & $\checkmark$ & $\checkmark$ & $\checkmark$ & $\checkmark$ \\
\bottomrule
\end{tabular}
\end{table}

\paragraph{Rating criteria.}
Each system is rated at the strongest configuration its public documentation commits to, uniformly across rows. \checkmark: natively satisfied in that configuration; $\triangle$: claimed but delivered only via an external substrate or an uncommitted deployment mode; $\times$: not established under our definition. A $\times$ is a statement about scope, not value: many cited systems are well-designed answers to neighboring problems. \emph{Bearer-token baseline} characterizes the prevalent agent-framework pattern (LangChain~\citep{langchain2024}, AutoGen~\citep{autogen2023}, CrewAI~\citep{crewai2024}, OpenAI Agents SDK~\citep{openai2025agentssdk}) rather than ranking specific frameworks. \emph{\sudp{}'s RFS} at the authority level additionally depends on the deployment honouring \S\ref{sec:rotation}'s rotation/revocation cadence at $E$.

\paragraph{Landscape and \sudp's position.}
The table shows two neighboring clusters. \emph{Secret managers and credential proxies} (Vault, Aegis, Claude Agent SDK proxy, IronClaw) keep reusable authority outside the requester and cover Secret Confidentiality to varying degrees, but issue no operation-bound \(U\)-rooted authorization. \emph{Agentic delegation protocols} (Authenticated Delegation, SAGA, OIDC-A, HDP, Agentic JWT) build authorization, scoping, and provenance for agent principals, some as OAuth/OIDC extensions and others as standalone token schemes, but leave reusable authority inside the requester boundary. \sudp{} natively satisfies AV, OB, UB, CRC, CSB, and RFS and is FI-positive; CMC holds only inside a hardware-rooted runtime.

\section{Secret-Use Delegation Protocol}
\label{sec:protocol}

\sudp{} is a protocol-level answer to capability-constrained secret use. The unit of delegation is the \emph{use} of a secret for one specific authorized operation $o$, not the secret itself: the requester is delegated the right to \emph{cause} an authorized use, the custodian is delegated the right to \emph{perform} the use within \(U\)'s signed bounds, and reusable authority never crosses the requester boundary. This section fixes \sudp's roles and trust model (\S\ref{sec:roles}), then defines \sudp{} as an abstract protocol over generic cryptographic interfaces (\S\ref{sec:keyhierarchy}--\ref{sec:phase3}) together with the rotation discipline that delivers per-credential forward secrecy (\S\ref{sec:rotation}).

\subsection{Roles and Trust Model}
\label{sec:roles}

\sudp{} solves the Agent Secret Use problem (\S\ref{sec:problem}) by introducing one new protocol role: a \emph{custodian} \(T\) that holds authority-bearing material as sealed state \(\Sigma\), issues freshness, verifies \(U\)-rooted operation-bound grants, and performs the authorized secret-backed action within the bounds \(U\) signed. The custodian is a protocol role, not necessarily a deployment component: realizations include a local proxy, a hosted proxy, an OS-level credential mediator, an HSM- or TEE-backed service, or, in cooperative settings, the environment itself. The agentic specialization (how \(R\) is drawn around the LLM-context exposure surface and how native calls are bound to canonical operations) is deferred to \S\ref{sec:interface}.

\paragraph{\sudp{} entities.}
Three parties run the protocol:
\begin{itemize}[leftmargin=*,topsep=2pt,itemsep=0pt]
  \item \(U\): the \emph{authorizer}, governing whether a secret-backed operation may proceed.
  \item \(R\): the \emph{requester}, initiating an operation request, typically an agent.
  \item \(T\): the \emph{custodian}, holding the sealed state \(\Sigma\), validating authorization, redeeming operation-bound grants, and using protected authority material to perform the secret-backed operation; distinct from \(U\).
\end{itemize}
We say \emph{authorizer} rather than \emph{user} because the role is functional, not human: \(U\) is typically a person signing on a personal device, but may equally be an agent holding further-rooted authority in a chained delegation. The environment \(E\) is as in Agent Secret Use: the execution target, outside the protocol, assumed only to accept its existing credential interface.

\paragraph{\sudp{}-specific trust assumptions.}
Beyond authorizer-controlled authorization (\S\ref{sec:trust-model}), \sudp{} adds two assumptions. \emph{Restricted trust in the custodian}: \(T\) is trusted only to validate authorization, redeem grants, and transiently use secret material during valid consumption; \(R\) is never trusted with secret access. \emph{Sealed persistent state}: \(\Sigma\) alone does not suffice for standalone recovery, and any authorized extraction leaves \(T\) only in recipient-protected form.

\subsection{Key Hierarchy}
\label{sec:keyhierarchy}

\sudp{} uses the standard envelope-encryption pattern (Figure~\ref{fig:hierarchy}). The protected state \(M\) (the contents of \(\Sigma\) the protocol secures) is sealed once under a state-encryption key \(K\); \(K\) is in turn wrapped separately under each authorizer credential's wrapping key \(W_c\). This separation is deliberate: rotating \(K\) does not require interacting with every credential (each \(W_c\) simply rewraps the new \(K\)), and registering a new credential does not require re-encrypting \(M\) (only a new \([K]_c\) is appended). Formally,
\[
  C \gets \mathsf{Enc}_K(M), \qquad M \gets \mathsf{Dec}_K(C),
\]
and for each enrolled credential \(c\),
\[
  [K]_c \gets \mathsf{Wrap}_{W_c}(K), \qquad K \gets \mathsf{Unwrap}_{W_c}([K]_c).
\]
The persistent cryptographic state consists of \(C\) and \(\{[K]_c\}\); \(\{W_c\}\) defines the recoverability structure of \(K\). By construction, the requester never receives \(K\), any \(W_c\), or any value from which either can be efficiently derived.

\begin{figure}[t]
\centering
\resizebox{0.95\columnwidth}{!}{%
\begin{tikzpicture}[
  font=\footnotesize,
  >=Stealth,
  box/.style={draw, rounded corners, minimum height=8mm, minimum width=46mm, align=center, fill=white},
  role/.style={font=\scriptsize, align=left, anchor=west},
  zone/.style={font=\scriptsize\bfseries, align=left, anchor=west},
  edge/.style={->, thick},
  legendbox/.style={draw, rounded corners, minimum height=3mm, minimum width=4mm, inner sep=0pt}
]
  \node[box, fill=blue!8] (auth) {$\mathit{Aut}_c$: $sk_c$, PRF key};
  \node[box, below=4mm of auth, fill=blue!8] (y) {$y_c \gets \mathsf{PRF}_c(\eta_c)$};
  \node[box, below=4mm of y, fill=gray!12] (w) {$W_c \gets \mathsf{KDF}(y_c;\, \eta_c,\, \mathsf{DS}_{\mathsf{wrap}}\,\|\,\mathit{cid}_c\,\|\,\mathit{ver})$};
  \node[box, below=4mm of w, fill=red!7] (k) {$[K]_c \gets \mathsf{Wrap}_{W_c}(K)$};
  \node[box, below=4mm of k, fill=red!7] (m) {$C \gets \mathsf{Enc}_K(M)$};

  \draw[edge] (auth) -- (y);
  \draw[edge] (y) -- (w);
  \draw[edge] (w) -- (k);
  \draw[edge] (k) -- (m);

  \node[role] at ($(auth.east) + (0.3, 0)$) {authorizer credential};
  \node[role] at ($(y.east) + (0.3, 0)$) {credential-scoped PRF output};
  \node[role] at ($(w.east) + (0.3, 0)$) {per-credential wrapping key (transit \(U\!\to\!T\))};
  \node[role] at ($(k.east) + (0.3, 0)$) {wrapped state key};
  \node[role] at ($(m.east) + (0.3, 0)$) {encrypted protected state};

  \begin{scope}[on background layer]
    \node[draw, dashed, rounded corners, inner sep=2mm, fit=(auth)(y), fill=blue!3] (zA) {};
    \node[draw, dashed, rounded corners, inner sep=2mm, fit=(w), fill=gray!4] (zB) {};
    \node[draw, dashed, rounded corners, inner sep=2mm, fit=(k)(m), fill=red!3] (zC) {};
  \end{scope}

  \node[zone, rotate=90, anchor=south] at ($(zA.west) + (-0.3, 0)$) {Authorizer-held credential};
  \node[zone, rotate=90, anchor=south] at ($(zB.west) + (-0.3, 0)$) {Transit};
  \node[zone, rotate=90, anchor=south] at ($(zC.west) + (-0.3, 0)$) {Custodian-held sealed state};

  \node[anchor=north west, font=\scriptsize] (lg) at ($(zC.south west) + (-0.3, -0.3)$)
    {\begin{tabular}{@{}l@{\hspace{4pt}}l@{\hspace{10pt}}l@{\hspace{4pt}}l@{\hspace{10pt}}l@{\hspace{4pt}}l@{}}
      \tikz\node[legendbox,fill=blue!8]{}; & authorizer secret / non-extractable &
      \tikz\node[legendbox,fill=gray!12]{}; & transit (confidential) &
      \tikz\node[legendbox,fill=red!7]{}; & persistent sealed state
    \end{tabular}};
\end{tikzpicture}}
\caption{\sudp{} key hierarchy, organized in three trust zones. \emph{Authorizer-held credential} material (blue) is held on the authorizer side: $sk_c$ and the PRF key never leave $\mathit{Aut}_c$, and the PRF output $y_c$ is released only within $U$'s trusted client. \emph{Transit} (gray) is the wrapping key $W_c$ that $U$ derives from $y_c$ and sends to $T$ under confidentiality. \emph{Custodian-held sealed state} (red) is the persistent wrapping/encryption chain stored in $\Sigma$: $[K]_c$ and $C$. Arrows trace the unlock chain downward; the figure shows one credential branch, with the acting-credential instance denoted $W^\star$ in Phase~II (\S\ref{sec:phase2}).}
\label{fig:hierarchy}
\end{figure}

\subsection{Authorized Operation}
\label{sec:opgrant}

Rather than authorizing decryption in the abstract, \sudp{} authorizes a specific secret-backed operation together with its redemption binding and validity conditions. A well-formed authorization object encodes three aspects: (i)~\emph{authorization semantics}: what is approved; (ii)~\emph{redemption binding}: who may redeem and, if extracting, who receives; and (iii)~\emph{validity constraints}: how long the authorization is valid and how it resists replay and substitution. Each component corresponds to a distinct security obligation: the semantics must be reviewable at $U$, the binding must resist substitution, and the validity must resist replay. Freshness is supplied by the Phase~II freshness token $r$ (\S\ref{sec:phase2}) rather than carried inside $o$; $o$ commits to $r$ implicitly through the Phase~II binding.

\begin{definition}[Authorized operation]\label{def:op}
An \emph{authorized operation} is a canonical protocol tuple
\[
  o := (\mathsf{act},\, \mathsf{bind},\, \mathsf{valid}),
\]
where \(\mathsf{act} := (\mathsf{type},\mathsf{target},\mathsf{scope})\), \(\mathsf{bind} := (\mathsf{redeemer},\mathsf{recipient})\), and \(\mathsf{valid} := (\mathsf{expiry},\mathsf{multiplicity})\). Here \(\mathsf{type}\) specifies the semantic class of the secret-backed action (\eg, \(\mathsf{use}\), \(\mathsf{export}\), \(\mathsf{write}\), \(\mathsf{rotate}\), \(\mathsf{enroll}\), \(\mathsf{revoke}\)); \(\mathsf{target}\) identifies the protected object; \(\mathsf{scope}\) carries canonicalized operation-specific constraints; \(\mathsf{redeemer}\) identifies the party entitled to redeem; \(\mathsf{recipient}\) identifies the intended recipient of any extracting delivery; \(\mathsf{expiry}\) bounds the validity window; and \(\mathsf{multiplicity} \in \mathbb{Z}^{+} \cup \{\infty\}\) sets the grant's execution budget: \(T\) admits at most \(\mathsf{multiplicity}\) Phase~III executions under the grant before \(\mathsf{expiry}\); \(1\) consumes the grant at first redemption (high-authority transaction-style actions), and \(\infty\) yields a \emph{standing grant} (\(U\)-declared low-risk repeat patterns such as inbox reads). Freshness for the Phase~II verification is enforced separately, via the single-use token $r$ issued at Phase~II.1 and consumed at Phase~II.3.
\end{definition}

We reserve specific terms: a \emph{grant} \(G\) is the protocol artifact representing \(U\)'s authorization of \(o\); it commits to \(o\) and conveys for redemption both the authorization signature and the acting credential's wrapping key (the full structure is in \S\ref{sec:phase2}). The wrapping key is secret: \(G\) is therefore not a public bearer token, and the \(U \to T\) leg carrying it requires end-to-end confidentiality. A \emph{redeemed grant} \(\rho\) is \(G\)'s custodian-internal post-validation form. \sudp{} binds \(G\) and \(\rho\) to the same \(\beta\) commitment in Phase~II so that the artifact \(T\) redeems matches the one \(U\) authorized.

\sudp's cryptographic binding attaches to the canonical bytes of \(o\). Two deployment obligations anchor those bytes at the endpoints of the authorization; AV and OB hold when both are met.

\paragraph{Render at \(U\).}
\(\mathsf{Render}(o)\) is deterministic, complete, and confined to \(U\)'s trusted client, so \(U\) approves exactly what \(\mathsf{H}(o)\) commits.

\paragraph{Execution at \(E\).}
The action performed at \(E\) is a deterministic function of \(o\), the material \(T\) unwraps, and any auxiliary content that \(o\) commits by hash. No authority-relevant input enters between \(U\)'s authorization and the action.

\subsection{Cryptographic Primitives}
\label{sec:primitives}

We fix primitive interfaces here; concrete algorithm choices appear in \S\ref{sec:concrete}.
\begin{itemize}[leftmargin=*,topsep=2pt,itemsep=1pt]
  \item \(\mathsf{H}\): collision-resistant hash.
  \item \(\mathsf{KDF}(\mathit{ikm};\,\mathit{salt},\,\mathit{info})\): HKDF-style extract-then-expand with explicit domain separation in \(\mathit{info}\).
  \item \((\mathsf{Enc},\mathsf{Dec})\), \(c \gets \mathsf{Enc}_k(m;\,\mathit{ad})\): IND-CCA AEAD with associated-data authentication.
  \item \((\mathsf{Sig},\mathsf{Vrfy})\): EUF-CMA secure.
  \item \((\mathsf{Encap},\mathsf{Decap})\): IND-CCA2 secure.
  \item \(\mathsf{CSPRNG}\): cryptographically secure randomness source.
  \item \((\mathsf{Wrap},\mathsf{Unwrap})\), \(E \gets \mathsf{Wrap}_W(K)\): key-wrap interface specialized to key material.
  \item \(\mathsf{PRF}_c\): authenticator-bound pseudo-random function, invoked only inside the module \(\mathit{Aut}_c\) (specified below).
\end{itemize}
The \(\mathsf{Wrap}\) interface binds wrapping context through the derivation of \(W\) rather than through a separate parameter: because \(W\) is produced by a domain-separated \(\mathsf{KDF}\) keyed to the credential, salt, and version (\S\ref{sec:keyhierarchy}), any attempt to unwrap under a \(W\) derived from different context material fails by construction, and the interface accommodates both associated-data-free wrap constructions (such as AES-KW/KWP) and AEAD-as-wrap profiles that additionally authenticate the same domain-separation labels.

\paragraph{Authenticator module \(\mathit{Aut}_c\).}
Each authorizer-side credential \(c\) is modeled as a tamper-resistant module \(\mathit{Aut}_c\) with non-extractable internal keys. Under a single \(U\)-verified invocation, \(\mathit{Aut}_c\) produces one signature \(\sigma \gets \mathsf{Sig}_{\mathit{sk}_c}(\mu)\) over an externally supplied challenge \(\mu\), and evaluates \(\mathsf{PRF}_c\) on one or more caller-supplied salts \(\eta_i\), returning the outputs \(y_i \gets \mathsf{PRF}_c(\eta_i)\). The number of PRF evaluations admissible per invocation is a parameter of \(\mathit{Aut}_c\)'s interface; the rotation-class operations of \S\ref{sec:phase3} require at least two. The signing key and the PRF key never leave the module. Enrollment releases public material \((\mathit{cid}_c, \mathit{pk}_c)\). Every \(\mathit{info}\) or \(\mathit{ad}\) argument carrying a domain-separation label \(\mathsf{DS}_\star\) uses a stable, context-disjoint string; labels are pairwise disjoint and no label is ever reused. The protocol equations explicitly name \(\mathsf{DS}_{\mathsf{bind}}\) and \(\mathsf{DS}_{\mathsf{wrap}}\); concrete profiles may use additional disjoint labels for sealing, delivery, and transport without affecting the abstract contract.

\paragraph{Canonical serialization.}
\(\mathsf{H}(o)\) presumes a deterministic encoding of the structured value \(o\) that is also identical at \(U\) and \(T\). A concrete profile must fix this serialization (for instance, a canonical CBOR or RFC~8785 JCS encoding of the \((\mathsf{act}, \mathsf{bind}, \mathsf{valid})\) fields) and enforce it at both endpoints. Ambiguity in the encoding can defeat operation binding without violating collision resistance of \(\mathsf{H}\): two semantically distinct descriptors may serialize to the same bytes (so that an adversary substitutes \(o' \neq o\) with the same \(\mathsf{H}\)-input), or the same descriptor may serialize differently at \(U\) and \(T\) (so that \(U\) authorizes an operation whose canonical form at \(T\) differs from the one \(\mathsf{Render}\) presented at authorization time). Both failures collapse OB regardless of \(\mathsf{H}\)'s strength. The specific serialization is a profile parameter, not a protocol choice.

Pairwise communication runs over a pre-established authenticated channel; transport is assumed, not re-specified. Confidentiality is required on the \(U \to T\) leg (which carries \(W^\star\)) and on the final response leg (which may carry the delivery artifact \(\pi\) or a use-result \(\rho_{\mathsf{out}}\)). This requirement is on the logical \(U \to T\) leg end-to-end and is not in general satisfied by hop-by-hop transport security; a deployment with a plaintext-inspecting intermediary between \(U\) and \(T\) must supply end-to-end confidentiality at the application layer (\S\ref{sec:concrete-xdevice}). Integrity is required on every leg. The \(T \to U\) conveyance (Phase~II.1) carries only public material \((o, r, \{(\mathit{cid}_c, \eta_c)\})\) and does not require confidentiality; its only protection need is against denial-of-service tampering, which is detected at redemption through the signature on \(\beta\).

\subsection{Phase I: Setup}
\label{sec:phase1}

\begin{figure*}[!t]
\centering
\resizebox{0.75\textwidth}{!}{%
\begin{tikzpicture}[
  xscale=1.35,
  font=\footnotesize,
  >=Stealth,
  header/.style={draw, rounded corners, minimum height=5mm, minimum width=22mm, align=center, font=\small\bfseries, fill=white},
  lifeline/.style={draw, thick, gray!45, dash pattern=on 1.2pt off 1.2pt},
  msg/.style={->, thick, >=Stealth},
  internal/.style={draw, rounded corners, fill=blue!6, inner sep=3pt, align=left, font=\scriptsize},
  branch/.style={draw, rounded corners, fill=blue!3, inner sep=1.5pt, align=center, font=\scriptsize, minimum width=22mm, minimum height=4.5mm},
  lbl/.style={font=\scriptsize, inner sep=1pt, fill=white},
  phase/.style={font=\scriptsize\bfseries, anchor=east, text=gray!55!black},
  output/.style={draw, rounded corners, dashed, fill=white, inner sep=1.5pt, align=center, font=\scriptsize}
]
  \node[header, fill=red!4]  (R) at (0, 0)    {Requester $R$};
  \node[header, fill=blue!6] (U) at (4.0, 0)  {Authorizer $U$};
  \node[header, fill=gray!8] (P) at (8.0, 0)  {Custodian $T$};
  \node[header]              (E) at (12.0, 0) {Env $E$};
  \foreach \x in {0, 4.0, 8.0, 12.0} {
    \draw[lifeline] (\x, -0.35) -- (\x, -8.0);
  }

  \node[phase] at (-1.0,-0.7) {II.1};
  \draw[msg] (0, -0.7) -- node[lbl] {request authorization for $o$} (8.0, -0.7);
  \draw[msg] (8.0, -1.2) -- node[lbl] {$(o,\, r,\, \{(\mathit{cid}_c, \eta_c)\})$} (4.0, -1.2);

  \node[phase] at (-1.0,-2.7) {II.2};
  \node[internal, fill=blue!10, minimum width=58mm] (ua) at (4.0, -2.7) {%
    $\beta := \mathsf{H}(\mathsf{DS}_{\mathsf{bind}}\,\|\,r\,\|\,\mathsf{H}(o))$;\\[1pt]
    invoke $\mathit{Aut}$; with responding credential $c^\star$:\\[1pt]
    \quad $\sigma^\star \gets \mathsf{Sig}_{\mathit{sk}_{c^\star}}(\beta)$\\[1pt]
    \quad $W^\star \gets \mathsf{KDF}(\mathsf{PRF}_{c^\star}(\eta_{c^\star});\,\eta_{c^\star},\,\mathsf{DS}_{\mathsf{wrap}}\|\mathit{cid}_{c^\star}\|\mathit{ver})$
  };
  \draw[msg] (ua.east) -- node[lbl, above=2pt, align=center, pos=0.5] {$(o,\, r,\, \mathit{cid}_{c^\star},$\\[-1pt]$W^\star,\, \sigma^\star,\, \mathit{opt})$} (8.0, -2.7);

  \node[phase] at (-1.0,-4.0) {II.3};
  \node[internal, minimum width=58mm] (p23) at (8.0, -4.0) {%
    consume $r$ from $\mathsf{S}$;\\[1pt]
    recompute $\beta'$, verify $\mathsf{Vrfy}_{\mathit{pk}_{c^\star}}(\beta',\sigma^\star) = 1$;\\[1pt]
    confirm $o.\mathsf{bind}.\mathsf{redeemer} = T_{\mathsf{id}}$, expiry
  };

  \node[phase] at (-1.0,-5.3) {Phase~III};
  \node[internal, minimum width=58mm] (p30) at (8.0, -5.3) {%
    $K \gets \mathsf{Unwrap}_{W^\star}([K]_{c^\star})$;\ $M \gets \mathsf{Dec}_K(C)$;\\[1pt]
    $s_o \gets M[o.\mathsf{act}.\mathsf{target}]$
  };

  \node[branch] (b1) at (8.0, -6.4) {III.1 \textsf{use}};
  \node[branch] (b2) at (8.0, -7.1) {III.2 \textsf{export}};
  \node[branch] (b3) at (8.0, -7.8) {III.3 \textsf{lifecycle}};
  \draw[->, thick] (p30.south) -- (b1.north);

  \draw[msg] (b1.east) -- node[lbl] {$\mathsf{execute}(o;\, s_o)$} (12.0, -6.4);
  \draw[->, thick, dashed] (b1.west) -- node[lbl] {result $\rho_{\mathsf{out}}$} (0, -6.4);

  \draw[->, thick, dashed] (b2.east) -- node[lbl, fill=none, above=0.5pt, pos=0.55] {$\pi = \mathsf{KEMSeal}(\mathit{pk}_{\mathsf{rcp}},\, s_o)$} node[lbl, fill=none, below=0.5pt] {to $o.\mathsf{bind}.\mathsf{recipient}$} (11.4, -7.1);

  \node[anchor=west, font=\scriptsize\itshape] at (b3.east) {\quad update sealed state $\Sigma$};
\end{tikzpicture}%
}

\caption{End-to-end \sudp{} flow across requester $R$, authorizer $U$, custodian $T$, and environment $E$. Phase II issues a $U$-signed grant $G$ committing to $o$; Phase III unwraps $K$ with $W^\star$ and dispatches on $o.\mathsf{act}.\mathsf{type}$ to \textsf{use}, \textsf{export}, or \textsf{lifecycle}. The $U \to T$ leg carries $W^\star$ under confidentiality. Solid arrows carry messages, dashed arrows return results, and shaded boxes are internal computation.}
\label{fig:flow}
\end{figure*}

Setup initializes the persistent cryptographic state required by the remaining phases. Its structural outcome is that, after the setup phase completes and transient values are erased, no persistent state held by any single party contains both \(K\) and any wrapping key \(W_c\), and every enrolled credential is on an independent authorization path to \(K\). The requester \(R\) is not involved.

\paragraph{I.1 Credential enrollment.}
\(U\) registers one or more authenticator credentials with \(T\). Each credential creation generates \(\mathit{Aut}_c\)'s internal PRF key and signing key (neither of which ever leaves the module) and releases a credential identifier \(\mathit{cid}_c\) together with the public verification key \(\mathit{pk}_c\):
\[
  (\mathit{cid}_c, \mathit{pk}_c) \gets \mathsf{Enroll}(\mathit{Aut}_c).
\]
\(T\) records the registration map \(\mathsf{Reg} := \{\mathit{cid}_c \mapsto \mathit{pk}_c\}_{c \in \mathsf{Creds}}\), which Phase~II.3 will use to verify authorization evidence.

\paragraph{I.2 Protected-state sealing.}
\(T\) generates a fresh state-encryption key and seals the initial protected state:
\[
  K \stackrel{\$}{\gets}\mathsf{CSPRNG},\qquad
  C \gets \mathsf{Enc}_K(M_0;\,\mathit{ver}).
\]
For each enrolled credential \(c\), \(U\) and \(T\) jointly derive a credential-specific wrapping key through a single \(U\)-verified invocation of \(\mathit{Aut}_c\). The authenticator evaluates its PRF under a freshly sampled salt \(\eta_c\) to produce \(y_c\); \(U\) derives the wrapping key \(W_c\) and transmits it to \(T\), which produces the wrapped entry \([K]_c\):
\begin{align*}
  \eta_c &\stackrel{\$}{\gets}\mathsf{CSPRNG},\quad y_c \gets \mathsf{PRF}_c(\eta_c),\\
  W_c &\gets \mathsf{KDF}(y_c;\,\eta_c,\,\mathsf{DS}_{\mathsf{wrap}}\,\|\,\mathit{cid}_c\,\|\,\mathit{ver}),\\
  [K]_c &\gets \mathsf{Wrap}_{W_c}(K).
\end{align*}
The authenticator-internal PRF key never leaves \(\mathit{Aut}_c\); only \(W_c\) crosses the \(U\!\to\!T\) channel, and is zeroized at \(T\) once \([K]_c\) is persisted. The domain-separation info binds \(\mathit{cid}_c\) and \(\mathit{ver}\), so derivations for distinct credentials and distinct wrapping epochs remain independent even when the same authenticator hosts several credentials.

\paragraph{I.3 Freshness state.}
\(T\) initializes a bounded single-use pool \(\mathsf{S}\) of freshness tokens \((r, \tau_r)\). Each token is issued and consumed exactly once in Phase~II and has short lifetime \(\tau_r\) (on the order of minutes), which bounds the set of outstanding authorizations.

\paragraph{I.4 Output.}
Phase~I leaves the persistent state
\[
  \Sigma_0 := \bigl(C,\;\{(\mathit{cid}_c,\eta_c,[K]_c)\}_{c\in\mathsf{Creds}},\;\mathsf{Reg},\;\mathit{ver}\bigr).
\]
Two invariants close the phase. \emph{Persistent-state separation:} after transient values from I.2 are zeroized, no persistent state held by any party contains both \(K\) and any \(W_c\); \(U\)'s client stores no extractable persistent key material, the authenticator retains non-extractable credential keys, and \(\Sigma_0\) alone is insufficient for plaintext recovery of \(M_0\). \emph{Uniform recoverability:} every enrolled credential \(c\) can reach \(K\) via one \(U\)-verified invocation of \(\mathit{Aut}_c\) together with \(\Sigma_0\) alone.

\subsection{Phase II: Authorization Grant}
\label{sec:phase2}

The grant phase converts an authorization decision into a \(U\)-signed grant \(G\) that commits cryptographically to \(o\); \(T\)'s post-validation record of \(G\) is the redeemed grant \(\rho\) (Figure~\ref{fig:flow}).

\paragraph{II.1 Grant request.}
\(R\) initiates by requesting authorization for \(o = (\mathsf{act}, \mathsf{bind}, \mathsf{valid})\), with \(\mathsf{bind}.\mathsf{redeemer} = T\) and, for extracting operations, \(\mathsf{bind}.\mathsf{recipient}\) set to the intended recipient's public key. The descriptor \(o\) is the cryptographic contract between \(T\) and \(U\): \(T\) binds to \(\mathsf{H}(o)\) in the channel binding \(\beta\), and \(U\) reviews \(\mathsf{Render}(o)\) before signing. \(T\) issues a fresh freshness token \(r \stackrel{\$}{\gets} \mathsf{CSPRNG}\), stores \((r, \tau_r)\) in \(\mathsf{S}\), and prepares the conveyance payload \((o, r, \{(\mathit{cid}_c, \eta_c)\}_{c \in \mathsf{Creds}})\). The \((\mathit{cid}_c, \eta_c)\) entries are public; \(U\) needs them to invoke the correct authenticator. The \(T \to U\) conveyance is a logical authenticated channel; its realization is a deployment dimension (\S\ref{sec:interface}). Confidentiality is not required on this leg: \(o\) is not sensitive, and any in-transit substitution will be detected at redemption against the signature on \(\beta\) (II.4). In direct execution (\(R = U\)), the conveyance collapses into the same session.

\paragraph{II.2 Authorizer signing.}
\(U\) renders \(o\) inside its trusted boundary and decides whether to approve. On approval, \(U\) selects an acting credential \(c^\star \in \mathsf{Creds}\) and, in a single \(U\)-verified invocation of \(\mathit{Aut}_{c^\star}\), produces a signature over \(\beta\) and a PRF-derived value for \(\eta_{c^\star}\).

\emph{Operation- and freshness-bound assertion.} \(U\) computes the binding
\[
  \beta := \mathsf{H}(\mathsf{DS}_{\mathsf{bind}}\,\|\,r\,\|\,\mathsf{H}(o)),
\]
passes \(\beta\) to \(\mathit{Aut}_{c^\star}\) as the signing challenge, and obtains
\[
  \sigma^\star \gets \mathsf{Sig}_{\mathit{sk}_{c^\star}}(\beta).
\]
The binding \(\beta\) commits to the session (via \(r\)), the domain-separated context label, and the operation (via \(\mathsf{H}(o)\)): any change to \(o\) between II.2 and II.3 causes the \(\beta'\) recomputed at redemption to diverge from the \(\beta\) signed into \(\sigma^\star\).

\emph{Derived wrapping material.} The same invocation evaluates \(\mathit{Aut}_{c^\star}\)'s PRF and derives the per-credential wrapping key for \(c^\star\):
\[
\begin{aligned}
  y^\star &:= \mathsf{PRF}_{c^\star}(\eta_{c^\star}),\\
  W^\star &\gets \mathsf{KDF}(y^\star;\,\eta_{c^\star},\,\mathsf{DS}_{\mathsf{wrap}}\,\|\,\mathit{cid}_{c^\star}\,\|\,\mathit{ver}).
\end{aligned}
\]
For rotation-class operations (\S\ref{sec:phase3}), \(U\) additionally samples a fresh next-salt \(\eta^{\mathsf{next}}_{c^\star}\), places it inside \(o.\mathsf{act}.\mathsf{scope}\) so that \(\beta\) commits to it via \(\mathsf{H}(o)\), and derives \(W^\star_{\mathsf{next}}\) analogously from \(\mathsf{PRF}_{c^\star}(\eta^{\mathsf{next}}_{c^\star})\) in the same invocation (\(\mathit{Aut}_{c^\star}\) admits two PRF evaluations per invocation, \S\ref{sec:primitives}).

\(U\) transmits the grant
\[
  G := (o,\; r,\; \mathit{cid}_{c^\star},\; W^\star,\; \sigma^\star,\; \mathit{opt}),
\]
to \(T\) over an end-to-end authenticated and confidential channel; here \(\mathit{opt} = (W^\star_{\mathsf{next}})\) appears only for rotation-class \(o\). Channel confidentiality is required end-to-end on this leg: any party that observes \(W^\star\) can unwrap \([K]_{c^\star}\), and the confidentiality of \(W^\star\) is not protected by the signature. The realization may be a direct \(U \to T\) session or an opaque relay through any intermediary (including \(R\)) under the cross-device envelope of \S\ref{sec:concrete-xdevice}, which keeps \(W^\star\) hidden from the relay; the two carrier topologies are catalogued in \S\ref{sec:concrete-topologies}. The asymmetry between the \(T \to U\) conveyance (public payload, no confidentiality needed) and the \(U \to T\) transmission (carries \(W^\star\), end-to-end confidentiality required) is a security-critical part of the channel model.

\paragraph{II.3 Grant redemption.}
On receipt of \(G\), \(T\) validates the grant by:
\begin{enumerate}[leftmargin=*,topsep=2pt,itemsep=1pt,label=\arabic*.]
  \item consume \((r,\cdot)\) from \(\mathsf{S}\) (reject if absent or expired);
  \item check \(o.\mathsf{valid}.\mathsf{expiry}\), \(o.\mathsf{bind}.\mathsf{redeemer}\), and act-specific shape conditions (rotation-class operations require \(W^\star_{\mathsf{next}}\) in \(\mathit{opt}\) and \(\mathsf{multiplicity} = 1\); extracting operations require \(o.\mathsf{bind}.\mathsf{recipient}\));
  \item recompute \(\beta':=\mathsf{H}(\mathsf{DS}_{\mathsf{bind}}\,\|\,r\,\|\,\mathsf{H}(o))\) from the received \(o\);
  \item verify \(\mathsf{Vrfy}_{\mathsf{Reg}[\mathit{cid}_{c^\star}]}(\beta',\sigma^\star)=1\).
\end{enumerate}
\(r\) is consumed regardless of which step fails, preventing reuse of the same freshness token. As standard hygiene against timing side channels, the comparison inside \(\mathsf{Vrfy}\) is performed in constant time. On success, \(T\) emits the redeemed grant
\[
  \rho := (o,\; \mathit{cid}_{c^\star},\; W^\star,\; \mathit{opt}).
\]
\(\rho\) is custodian-internal: \(R\) never sees it. \(r\)'s single-use consumption at II.3 prevents the same \((o, r, \sigma^\star)\) triple from being re-redeemed; \(\rho\)'s Phase~III lifecycle is governed by \(o.\mathsf{valid}\) (\S\ref{sec:opgrant}).

\paragraph{II.4 Anti-substitution.}
Substitution, tampering, or injection of a modified \(o'\) between approval and redemption changes \(\beta'\) and fails verification at II.3 under EUF-CMA of \(\mathsf{Sig}\); together with the trusted rendering of \(\mathsf{Render}(o)\) inside \(U\)'s boundary, this ensures the \(o\) that \(T\) validates is the \(o\) that \(U\) reviewed.

\paragraph{Batch extension.}
The grant generalizes from a single operation to an ordered list \(\mathsf{ops} := (o_1, \ldots, o_n)\) authorized under one signature by substituting \(\mathsf{H}(\mathsf{ops})\) for \(\mathsf{H}(o)\) in the binding,
\[
  \beta := \mathsf{H}(\mathsf{DS}_{\mathsf{bind}}\,\|\,r\,\|\,\mathsf{H}(\mathsf{ops})),
\]
where \(\mathsf{H}(\mathsf{ops})\) is computed over a canonical concatenation of the \(o_i\). \(U\)'s trusted rendering obligation extends to \(\mathsf{Render}(\mathsf{ops})\) as a single decision surface, and \(T\) validates each \(o_i\) under the II.3 obligations. The single-operation case is the \(n=1\) instance; the cryptographic core, redemption sequence, and security properties are unchanged.

\subsection{Phase III: Consumption}
\label{sec:phase3}

Consumption executes \(o\) against the sealed persistent state under \(\rho\), in a form determined by \(o.\mathsf{act}.\mathsf{type}\). \sudp{} distinguishes three semantic cases of Phase~III (not deployment APIs): \emph{non-extracting use} (III.1) spends the secret inside \(T\) and releases only the authorized result form to \(R\); \emph{recipient-protected export} (III.2) releases protected material only to the recipient named in \(o.\mathsf{bind}\) and is ASU-preserving only when that recipient lies outside \(R\)'s trust boundary; and \emph{state-update / lifecycle} operations (III.3) mutate protected state and are coupled to rotation. All three begin with the same access step (Figure~\ref{fig:phase3}).

\begin{figure}[t]
\centering
\resizebox{0.90\columnwidth}{!}{%
\begin{tikzpicture}[
  font=\footnotesize,
  >=Stealth,
  node distance=5mm,
  block/.style={draw, rounded corners, align=center, minimum height=8mm, inner sep=3pt, fill=white},
  access/.style={block, fill=green!8, minimum width=46mm},
  branch/.style={block, fill=green!4, minimum width=30mm},
  outlbl/.style={draw, rounded corners, dashed, align=center, minimum width=30mm, inner sep=3pt, font=\scriptsize\itshape, fill=white},
  arr/.style={->, thick, >=Stealth},
  lbl/.style={font=\scriptsize, inner sep=1pt}
]
  \node[block, fill=yellow!12] (in) at (0, 2.6) {input: $\rho$, $\Sigma$, $o$};
  \node[access] (acc) at (0, 1.4) {III.0\ \ unwrap $K$;\ \ select $s_o := M[o.\mathsf{act}.\mathsf{target}]$};

  \node[branch] (b1) at (-4.4, 0)  {III.1\ \ \textsf{use}};
  \node[branch] (b2) at ( 0  , 0)  {III.2\ \ \textsf{export}};
  \node[branch] (b3) at ( 4.4, 0)  {III.3\ \ \textsf{lifecycle}};

  \node[outlbl] (o1) at (-4.4, -1.6) {$\rho_{\mathsf{out}} = \mathsf{Release}(o)$\\\scriptsize $T$ presents $s_o$ to $E$;\\\scriptsize $R$ gets only the response};
  \node[outlbl] (o2) at ( 0  , -1.6) {sealed $\pi$\\\scriptsize opens only under $\mathit{sk}_{\mathsf{rcp}}$};
  \node[outlbl] (o3) at ( 4.4, -1.6) {committed $\Sigma'$\\\scriptsize $K$ rotated; $\eta_{c^\star}$ advanced};

  \draw[arr] (in) -- (acc);
  \coordinate (bus) at ($(acc.south)+(0,-0.25)$);
  \draw[thick] (acc.south) -- (bus);
  \draw[arr] (bus) -| (b1);
  \draw[arr] (bus) -- (b2);
  \draw[arr] (bus) -| (b3);
  \draw[arr] (b1) -- (o1);
  \draw[arr] (b2) -- (o2);
  \draw[arr] (b3) -- (o3);

  \begin{scope}[on background layer]
    \node[draw, rounded corners, dashed, gray!60, inner sep=4mm,
          fit=(acc)(b1)(b2)(b3),
          label={[font=\scriptsize\itshape, gray!55!black, anchor=north east, inner sep=2pt]north east:custodian $T$ boundary}] {};
  \end{scope}
\end{tikzpicture}}
\caption{Phase~III dispatch inside the custodian \(T\) boundary. III.0 unwraps \(K\) and selects the authority-bearing service secret \(s_o := M[o.\mathsf{act}.\mathsf{target}]\); dispatch on \(o.\mathsf{act}.\mathsf{type}\) then selects the consumption mode. \emph{Use:} \(T\) may present \(s_o\) to \(E\)'s native interface but never to \(R\); \(R\) receives only \(\mathsf{Release}(o) = \rho_{\mathsf{out}}\). \emph{Export:} \(T\) emits a sealed delivery \(\pi\) that opens only under \(\mathit{sk}_{\mathsf{rcp}}\). \emph{Lifecycle:} \(T\) commits a new sealed state \(\Sigma'\) with rotated \(K\) and advanced authenticator-credential salt \(\eta_{c^\star}\). In all three modes, reusable authority never crosses the \(T\) boundary toward \(R\).}
\label{fig:phase3}
\end{figure}

\paragraph{III.0 Access to protected state.}
We write \(s_o := M[o.\mathsf{act}.\mathsf{target}]\) for the authority-bearing service secret selected by an accepted operation (e.g., the API key, OAuth token, or signing key referenced by \(o.\mathsf{act}.\mathsf{target}\)); \(s_o\) is the only secret \(T\) materializes for \(o\), and the symbol \(s\) (Definition~\ref{def:authority-bearing}) is its abstract counterpart. Given \(\rho\), \(T\) uses the acting credential's wrapping key \(W^\star\) carried in \(\rho\) to unwrap \(K\), decrypts \(M\), and selects \(s_o\):
\[
  K \gets \mathsf{Unwrap}_{W^\star}([K]_{c^\star}),\quad
  M \gets \mathsf{Dec}_K(C).
\]
The \(\mathsf{Unwrap}\) succeeds only under the \(W^\star\) bound to the matching salt, credential identifier, and version at Phase~I.2; the \(\mathsf{Dec}\) call binds its AEAD associated data to the protected-state context; any failure aborts the phase without releasing plaintext. \(\rho\)'s lifetime in \(\mathsf{S}\) is governed by \(o.\mathsf{valid}\) (\S\ref{sec:opgrant}). On success, \(T\) holds \(K\) and \(M\) within its trusted execution boundary and dispatches on \(o.\mathsf{act}.\mathsf{type}\).

\paragraph{III.1 Non-extracting consumption (\(\mathsf{type}=\mathsf{use}\)).}
\(T\) applies the authorized action to \(M\) inside its own boundary and never releases secret material from \(M\) to \(R\). We write \(\mathsf{Release}(o) = \rho_{\mathsf{out}}\) for the result form returned to \(R\). The protocol guarantees that \(\mathsf{Release}(o)\) contains no plaintext element of \(M\); the substantive content of \(\mathsf{Release}(o)\) is released under \(o.\mathsf{act}.\mathsf{scope}\), since the downstream environment's response may itself carry sensitive or authority-amplified data. For signing operations specifically, the scope must bind the exact message or transaction digest, the intended verifier/domain/audience, and the replay domain, so that a \(\mathsf{Release}(o)\) signature is not usable outside the authorized context.

\paragraph{III.2 Recipient-protected extracting delivery (\(\mathsf{type}=\mathsf{export}\)).}
Some operations inherently require extraction: migrating a secret between devices, provisioning a downstream service, or handing off to a specialized sub-agent. \sudp{} admits these under the constraint that the extracted value leaves \(T\) only in a form cryptographically protected for a named recipient. For ASU-preserving execution, a well-formed export operation requires \(\mathsf{bind}.\mathsf{recipient}\) to denote a party outside \(R\)'s trust boundary; if \(U\) authorizes an export with \(\mathsf{bind}.\mathsf{recipient} = R\), the operation is a deliberate ownership-transfer that falls outside \sudp's CRC and ASU non-disclosure guarantees and must be surfaced as such at authorization time. Given \(\mathit{pk}_{\mathsf{rcp}}\) from \(o.\mathsf{bind}\), \(T\) uses a key-encapsulation mechanism to produce an ephemeral encapsulated key, derives a delivery key bound to the specific operation via \(\mathsf{H}(o)\), and seals the target:
\begin{align*}
  (K_{\mathsf{d}},\mathit{ct}_{\mathsf{d}}) &\gets \mathsf{Encap}(\mathit{pk}_{\mathsf{rcp}}),\\
  k_{\mathsf{d}} &\gets \mathsf{KDF}(K_{\mathsf{d}};\,\bot,\,\mathsf{H}(o)),\\
  \delta &\gets \mathsf{Enc}_{k_{\mathsf{d}}}(s_o;\,\mathsf{H}(o)).
\end{align*}
\(T\) emits the delivery artifact \(\pi := (\mathit{ct}_{\mathsf{d}}, \delta)\); we abbreviate this sealing as \(\pi = \mathsf{KEMSeal}(\mathit{pk}_{\mathsf{rcp}},\, s_o)\). Only the holder of \(\mathit{sk}_{\mathsf{rcp}}\) can \(\mathsf{Decap}\) to recover \(K_{\mathsf{d}}\), rederive \(k_{\mathsf{d}}\), and open \(\delta\). Binding both the KDF info and the AEAD associated data to \(\mathsf{H}(o)\) prevents an adversary from substituting a captured \(\pi\) into the transcript of a different authorized operation: any such substitution would change \(\mathsf{H}(o)\) and fail the AEAD integrity check. When the recipient is a third party, \(R\) may act as a relay and gains no additional capability because \(\pi\) transits \(R\) only in sealed form.

\paragraph{III.3 State-update / lifecycle consumption (\(\mathsf{type}\in\{\mathsf{write},\mathsf{rotate},\mathsf{enroll},\mathsf{revoke}\}\)).}
Operations that mutate \(M\) (writes, pure \(K\)-rotations, and lifecycle events such as credential enrollment or revocation) are structurally coupled to key rotation. A single consumption applies \(o\) to \(M\) to obtain \(M'\), samples a fresh state key \(K'\) under which \(M'\) is re-sealed, and replaces the acting credential's wrapping material using \(W^\star_{\mathsf{next}}\) carried in \(\mathit{opt}\):
\begin{align*}
  K' &\stackrel{\$}{\gets}\mathsf{CSPRNG},\\
  C' &\gets \mathsf{Enc}_{K'}(M';\,\mathit{ver}),\\
  E'_{c^\star} &\gets \mathsf{Wrap}_{W^\star_{\mathsf{next}}}(K').
\end{align*}
The update must be atomic as a protocol invariant: either \((C, \{[K]_c\})\) is replaced by \((C', \{E'_c\})\) in its entirety or neither is written. As an implementation requirement for atomicity, the new wrapping material \(\{E'_c\}\) must be persisted durably before the old entries are retired; any crash between the two steps must leave the old \((C, \{[K]_c\})\) recoverable, never a state in which both sides have been partially overwritten. Violating this order leaves the only decryption path to \(K\) unreachable and produces an unrecoverable \(\Sigma\). Multi-path recoverability requires rewrapping \(K'\) under each \(W_c\) for \(c \neq c^\star\); the structure of this rewrap is part of the security argument and is addressed in \S\ref{sec:rotation}.

\subsection{Rotation and Forward Secrecy}
\label{sec:rotation}

Rotation is a \emph{protocol invariant} of \sudp{}: every \sudp{}-conformant deployment advances both the wrapping epoch on every state-update and, on a deployment-defined cadence, the underlying authority-bearing secret at $E$ together with revocation of the retired secret. The first part is enforced by Phase~III.3 below; the second part is mandated by the protocol's \(\mathsf{type}=\mathsf{rotate}\) operation class (Phase~III.3, \S\ref{sec:phase3}) and its compliance requirement that authority-bearing secrets never persist beyond their rotation epoch. This section specifies the wrapping-side mechanism; the authority-side cadence is a deployment policy parameter.

\paragraph{Rotation triggers.}
Every state-update consumption (III.3) generates a fresh \(K'\) and rotates the acting credential's salt \(\eta_{c^\star} \to \eta^{\mathsf{next}}_{c^\star}\). Two specialized rotation modes arise as sub-cases:
\begin{itemize}[leftmargin=*,topsep=2pt,itemsep=2pt]
  \item \textbf{Rewrap} preserves the authority-bearing plaintext entries in \(M\) but reseals state under a fresh \(K'\) and fresh acting-credential wrapping material. Its purpose is credential hygiene: advancing a credential's salt independently of write traffic, or migrating it to new domain-separation parameters. Formally, rewrap is a III.3 execution in which the non-peer contents of \(M\) are unchanged and the \(\mathsf{Peer}[\mathit{cid}_c]\) entry is updated as described below.
  \item \textbf{Full rotation} rotates \(K\) and the acting credential's \(W\), and rewraps all other credentials under the same \(K'\). Formally, this is the ordinary III.3 execution with \(o.\mathsf{act}.\mathsf{type} = \mathsf{rotate}\) and \(M' = M\).
\end{itemize}
Coupling rotation to every write tightens the wrapping-epoch forward-secrecy profile without introducing a new ceremony: any adversary who has transiently obtained pre-rotation key material is locked out of wrapped state produced after the next write. Authority-level forward secrecy additionally requires that retired authority-bearing secrets at \(E\) be revoked on the deployment's rotation cadence; without that revocation, compromise of the current epoch confers the same authority as compromise of any prior epoch and the wrapping-side forward secrecy is meaningless against \(E\).

\paragraph{Forward secrecy.}
Let \(\mathsf{rot}_{c^\star}(t)\) denote the most recent time \(t' \le t\) at which credential \(c^\star\) performed a state-update. The following claim is instantiated by \sudp's per-write discipline.

\begin{lemma}[Per-authenticator-credential wrapping forward secrecy]\label{cl:pcfs}
For any credential \(c^\star\) and any \(t' < \mathsf{rot}_{c^\star}(t)\), compromise of the tuple \((y_{c^\star}^{(t')}, W_{c^\star}^{(t')})\) does not enable recovery of \(K\) or \(M\) from \(\Sigma_t\), under the PRF security of \(\mathit{Aut}_{c^\star}\) and of the \(\mathsf{KDF}\), and the key-wrap security of \(\mathsf{Wrap}\).
\end{lemma}
\begin{proof}[Proof sketch]
At \(\mathsf{rot}_{c^\star}(t)\), the next-salt \(\eta^{\mathsf{next}}_{c^\star}\) is sampled inside the rotation invocation from a fresh \(\mathsf{CSPRNG}\) draw, independent of all prior salts. Under the PRF security of \(\mathit{Aut}_{c^\star}\), the PRF output on \(\eta^{\mathsf{next}}_{c^\star}\) is indistinguishable from random given any number of PRF outputs under distinct prior salts; hence the compromised \(y_{c^\star}^{(t')}\) carries no information about the post-rotation \(y_{c^\star}^{\mathsf{new}}\). The domain-separated \(\mathsf{KDF}\) propagates this indistinguishability to \(W_{c^\star}^{\mathsf{new}}\). Finally, the wrapped entry in \(\Sigma_t\) is under \(W_{c^\star}^{\mathsf{new}}\), so no efficient adversary holding only pre-rotation state can unwrap.
\end{proof}

Lemma~\ref{cl:pcfs} is per-credential and activates only after a credential rotates; the residual gap for dormant credentials is catalogued as L3 in \S\ref{sec:security-limits}.

\paragraph{Default rewrap: in-state peer map.}
In multi-credential \sudp{} deployments, Phase~III.3 by credential \(c^\star\) must rewrap \(K'\) under \(\{W_c\}_{c \neq c^\star}\). The default construction adopts an \emph{in-state peer map}
\[ \mathsf{Peer} := \{\mathit{cid}_c \mapsto W_c\}_{c \in \mathsf{Creds}} \;\subseteq\; M, \]
which III.0's decryption of \(C\) transiently exposes to \(T\). Under this policy, III.3 completes by rewrapping \(E'_c \gets \mathsf{Wrap}_{W_c}(K')\) for each \(c \neq c^\star\), updating \(\mathsf{Peer}[\mathit{cid}_{c^\star}] := W^\star_{\mathsf{new}}\), and sealing:
\[
  \Sigma':=(C',\{(\mathit{cid}_c,\eta'_c,E'_c)\}_{c\in\mathsf{Creds}},\mathsf{Reg},\mathit{ver}),
\]
with \(\eta'_{c^\star}=\eta^{\mathsf{next}}_{c^\star}\) and \(\eta'_c=\eta_c\) for \(c \neq c^\star\). \(\rho\) is marked consumed; as an implementation hardening note, \(T\) zeroizes \(K\), \(K'\), all \(W\) values, and \(M\) after persistence. The cross-credential exposure entailed by the in-state peer map is catalogued as L4 in \S\ref{sec:security-limits}; alternative rewrap policies (all-present rotation, append-only catch-up chain) are discussed in \S\ref{sec:design-choices}.

\paragraph{Lifecycle extensions: enrollment and revocation.}
Credential addition and removal are ordinary III.3 executions with \(o.\mathsf{act}\) naming the credential affected and \(M'\) encoding the change. \emph{Enrolling} a credential \(c^+\) proceeds in two coupled \(U\)-verified invocations bundled into one operation: an enrollment invocation of \(\mathit{Aut}_{c^+}\) that yields \((\mathit{cid}_{c^+}, \mathit{pk}_{c^+}, \eta_{c^+}, W_{c^+})\), and the acting-credential rotation invocation of \(\mathit{Aut}_{c^\star}\) that authorizes the write. \(T\) computes \([K]_{c^+} \gets \mathsf{Wrap}_{W_{c^+}}(K')\), adds \((\mathit{cid}_{c^+}, \eta_{c^+}, [K]_{c^+})\) to \(\Sigma'\), and updates \(\mathsf{Peer}[\mathit{cid}_{c^+}] := W_{c^+}\). \emph{Revocation} removes the target credential from \(\mathsf{Reg}\), from the wrapping set, and from \(\mathsf{Peer}\), and reseals \(M'\) under a fresh \(K'\). Neither operation requires machinery beyond what has already been specified.

\section{Secret-Use Interfaces for Agentic Systems}
\label{sec:interface}

Applying \sudp{} to an agent system requires identifying the trust boundaries and mapping the protocol roles accordingly. This section describes how \sudp{} is applied in an LLM-based agent system; Figure~\ref{fig:agentic-interface} illustrates the interface.

\begin{figure}[t]
\centering
\resizebox{0.90\columnwidth}{!}{%
\begin{tikzpicture}[
  font=\small,
  >=Stealth,
  msg/.style={->, thick, >=Stealth},
  msglbl/.style={font=\scriptsize, inner sep=2pt, align=center}
]
  \node[draw, rounded corners, minimum height=14mm, minimum width=48mm, align=center, fill=blue!5] (U) at (0, 3.2)
    {\textbf{Authorizer} \(U\) + authenticator \(\mathit{Aut}_c\)\\[1pt]\scriptsize\emph{reviews \(o\), signs \(\beta\)}};

  \node[draw, rounded corners, minimum height=22mm, minimum width=32mm, align=center, fill=green!4] (T) at (0, 0)
    {\textbf{Custodian} \(T\)\\[1pt]\scriptsize redeems \(G\), spends \(s\)\\[1pt]{\scriptsize\itshape sealed state \(\Sigma\)}};

  \node[draw, rounded corners=6pt, minimum height=14mm, minimum width=34mm, align=center, fill=gray!5, text=black!70] (Ext) at (5.2, 0)
    {\textbf{Environment} \(E\)\\[1pt]\scriptsize\emph{(outside protocol)}};

  \node[draw, rounded corners, fill=white, font=\scriptsize, align=center,
        minimum width=32mm, minimum height=8mm, inner sep=1pt] (Rtool) at (-4.9, 0.95)
    {tool runtime};
  \node[draw, rounded corners, fill=white, font=\scriptsize, align=center,
        minimum width=32mm, minimum height=8mm, inner sep=1pt] (Rllm) at (-4.9, -0.95)
    {LLM};
  \draw[->, semithick, gray!55] ([xshift=-5mm]Rllm.north) -- node[font=\tiny\itshape, left=-1pt, text=gray!55!black] {tool call} ([xshift=-5mm]Rtool.south);
  \draw[->, semithick, gray!55] ([xshift=5mm]Rtool.south) -- node[font=\tiny\itshape, right=-1pt, text=gray!55!black] {observation} ([xshift=5mm]Rllm.north);
  \begin{scope}[on background layer]
    \node[draw, rounded corners, fill=orange!5, inner sep=3mm, fit=(Rtool)(Rllm)] (R) {};
  \end{scope}
  \node[font=\small, anchor=south west, inner sep=1pt, align=left, xshift=2pt] at (R.north west)
    {\textbf{Requester \(R\)}\\[-1pt]{\scriptsize\itshape model-influenced boundary}};

  \draw[msg] (U.south) -- node[msglbl, right, xshift=2pt]
    {grant \(G\) {\scriptsize (cross-device)}} (T.north);

  \draw[msg] (Rtool.east) -- (Rtool.east -| T.west);
  \draw[msg] (Rllm.east)  -- (Rllm.east  -| T.west);
  \node[msglbl, fill=white, inner sep=2pt] at ($(R.east)!0.5!(T.west)$) {propose \(o\)};

  \coordinate (rdrop) at (0, -2.05);
  \draw[->, dashed, semithick, gray!60]
    (T.south) -- (rdrop) -- node[msglbl, below=1pt, midway, text=gray!50!black] {result \(\rho_{\mathsf{out}}\)} (R.south east |- rdrop) -- (R.south east);

  \draw[msg] (T.east) -- node[msglbl, above]
    {execute(\(o\); \(s\))} (Ext.west);
\end{tikzpicture}}
\caption{The agentic secret-use interface.}
\label{fig:agentic-interface}
\end{figure}

\subsection{The Agentic Requester Boundary}
\label{sec:interface-boundary}

In an agentic system, the agent's actions are driven by the output of the LLM, which is in turn shaped by whatever enters its context. Retrieved documents, tool outputs, and memory reads can carry injected instructions, and the model may hallucinate even on benign input; model output is thus untrusted, as is any component acting on it. Therefore, the requester boundary must enclose every component whose behavior model output can steer, including the LLM runtime, the tool scheduler and tool clients (MCP-style runtimes among them), and agent-visible memory. The custodian \(T\) and the authenticator \(\mathit{Aut}_c\) must accordingly remain outside this boundary; hosting either within model-influenced code would reinstate authorization by exposure.

\subsection{Where SUDP Applies in an Agentic System}
\label{sec:interface-layers}

A single criterion determines whether a call enters the \sudp{} path: its execution must exercise authority-bearing material (Definition~\ref{def:authority-bearing}) that \(U\) has enrolled at an external environment \(E\). Calls that exercise no enrolled authority, such as a public web search or a local file read, pass through unmodified. The criterion applies at both layers of an agent's secret use.

\paragraph{Tool-call layer.}
Tool calls reach external environments such as code hosts and payment rails, and meet the criterion whenever execution would present a credential enrolled by \(U\), for instance a mailbox token or a cloud key. These calls are the motivating workload of the agentic specialization; \S\ref{sec:interface-compile} describes their mediation.

\paragraph{LLM-API layer.}
The agent authenticates to its model provider with an API key. Once \(U\) enrolls that key, it routes through \sudp{} like any other credential; since these calls are frequent and uniform, the authorizer will typically cover them with a standing grant of broad scope, which preserves the confidentiality properties (CRC, CSB) and avoids a ceremony per call.

\paragraph{Scope.}
\sudp{} addresses authorization integrity and secret confidentiality. Behavioral safety, the question of which operations an agent chooses to propose, is a separate concern; runtime monitors, policy DSLs, taint tracking, and sandboxes constrain proposals and compose with \sudp{} (\S\ref{sec:positioning}). The protocol likewise fixes the authorization mechanism while leaving approval policy to the deployment. The authorizer chooses the scope, validity window, and execution budget of each grant, from single-use to standing (\S\ref{sec:opgrant}), and AV, OB, and UB hold at every granularity. Per-request approval is therefore the narrowest policy setting rather than a protocol obligation.

\subsection{From Native Calls to Canonical Operations}
\label{sec:interface-compile}

\(R\) needs no protocol integration. It issues the native call it would issue without the protocol, with an opaque \emph{credential reference} in place of the secret: the reference names an enrolled credential but confers no authority, and can therefore reside inside \(R\). At the mediation boundary the call is canonicalized into \(o\) (target credential, method, authority-relevant fields, body digest) and rendered to \(U\) (\S\ref{sec:opgrant}); after the grant, \(T\) executes the accepted call with the secret substituted for the reference. Mediation is thus transparent to the agent: any tool capable of issuing its native call participates unmodified. The completeness condition (P1 in \S\ref{sec:security}) fixes the canonicalization obligation: every field that can affect the authority exercised at \(E\) must be bound in \(o\).

\section{Standard Construction}
\label{sec:concrete}

This section gives the standards-based cryptographic instantiation of the protocol: each abstract interface of \S\ref{sec:protocol} is fixed to a standardized component. The construction introduces no new cryptographic assumption.

\subsection{Primitives}
\label{sec:concrete-primitives}

\paragraph{Authenticator \(\mathit{Aut}_c\).}
The abstract protocol requires a tamper-resistant module \(\mathit{Aut}_c\) that (i) signs a challenge under user verification and (ii) emits an authenticator-bound pseudo-random value scoped per credential. We instantiate \(\mathit{Aut}_c\) with WebAuthn credentials using the PRF extension~\citep{webauthn2019,fido2prf2023}.

The abstract \(\sigma \gets \mathsf{Sig}_{\mathit{sk}_c}(\beta)\) interface is realized by a WebAuthn \emph{assertion} produced with challenge \(=\beta\), using ES256/P-256 as the profile-selected COSE signing algorithm. The corresponding \(\mathsf{Vrfy}\) check covers the assertion signature together with the assertion's \texttt{rpId} hash, origin, user-verification flag, and credential-ID binding; a signature-verification success without these structural checks does not satisfy the \(\mathsf{Sig}\) interface in this profile.

The authenticator-bound pseudo-random value is realized by the WebAuthn PRF extension, backed by CTAP \texttt{hmac-secret} where available, with a per-credential salt; credentials or assertions whose response does not contain the PRF output are rejected by the profile. Two compliance implications follow: (a) binding PRF output to a passkey means deletion of the passkey implies loss of access to any data encrypted under the derived key; (b) per-credential PRF salts support domain separation across credentials.

\paragraph{Key derivation.}
The abstract protocol requires a KDF that supports extract-then-expand semantics and domain separation. We instantiate \(\mathsf{KDF}\) with HKDF-SHA-256~\citep{krawczyk2010hmac}, following NIST SP 800-56C key-derivation guidance~\citep{chen2018kdfrec}. Domain separation is achieved by encoding a disjoint \(\mathsf{DS}_\star\) label in the \texttt{info} argument of each derivation; this is sufficient to ensure that distinct derivations produce independent keys under the hash-function random-oracle idealization.

\paragraph{Sealing and key wrapping.}
\(\mathsf{Enc}/\mathsf{Dec}\) is instantiated with a standard AEAD scheme such as AES-GCM~\citep{dworkin2007nist} or ChaCha20-Poly1305~\citep{nir2018chacha}, including its extended-nonce XChaCha20 variant~\citep{arciszewski2020xchacha}; profiles must enforce nonce uniqueness/freshness required by the selected AEAD. The AEAD's associated-data input carries the \(\mathsf{DS}_\star\) label together with the version identifier \(\mathit{ver}\). For \((\mathsf{Wrap},\mathsf{Unwrap})\), \sudp{} binds wrapping context through the derivation of \(W\) rather than through an explicit parameter to \(\mathsf{Wrap}\). Because \(W_c \gets \mathsf{KDF}(y_c;\,\eta_c,\,\mathsf{DS}_{\mathsf{wrap}}\,\|\,\mathit{cid}_c\,\|\,\mathit{ver})\) already embeds the credential identifier, salt, and version (\S\ref{sec:keyhierarchy}), the interface can be instantiated directly as AES-KW/KWP~\citep{dworkin2012keywrap,schaad2002keywrap,housley2009kwp} wrapping \(K\) under \(W_c\) without associated data. Profiles that prefer AEAD-as-wrap may additionally authenticate the same domain-separation labels as associated data, yielding the same abstract contract.

\paragraph{Recipient-protected export.}
For Phase~III.2, the abstract protocol requires a KEM that admits a recipient-specific encapsulation. We instantiate \((\mathsf{Encap},\mathsf{Decap})\) with HPKE~\citep{barnes2022hpke} rather than an ECIES-style construction such as DHIES~\citep{abdalla2005dhies}: HPKE provides a standardized hybrid encryption interface with clear security properties and parameter choices, and it composes cleanly with HKDF-based domain-separation labels.

\paragraph{Transport.}
\label{sec:concrete-transport}
Pairwise communication runs over TLS~1.3~\citep{rescorla2018tls13}. TLS supplies the authenticated and confidential channel assumed by Phase~II on the \(U\!\to\!T\) leg, not a cryptographic layer we extend. Neither TLS session keys nor TLS peer identities enter the protocol's cryptographic commitments: all protocol-level binding is at the application layer. When \(U\) and \(T\) do not share a live TLS session, Phase~II's channel assumptions are realized by the cross-device HPKE profile of \S\ref{sec:concrete-xdevice}. The same profile applies end-to-end whenever a plaintext-inspecting intermediary, such as an L7 proxy or a TLS-terminating load balancer, sits between \(U\) and \(T\), since hop-by-hop TLS does not realize the \(U \to T\) confidentiality assumption (\S\ref{sec:primitives}).

The grant \(G\) itself is a sender-constrained authorization token in the sense of DPoP~\citep{fett2023dpop}, with \(\sigma^\star\) binding operation and freshness to the acting credential. Table~\ref{tab:profile} maps each abstract cryptographic primitive of \S\ref{sec:primitives} to its concrete realization.

\begin{table}[t]
\centering
\footnotesize
\caption{\sudp{} cryptographic profile: abstract primitive interfaces and their standard-component instantiations.}
\label{tab:profile}
\begin{tabular}{@{}>{\raggedright\arraybackslash}p{0.32\columnwidth}>{\raggedright\arraybackslash}p{0.42\columnwidth}>{\raggedright\arraybackslash}p{0.14\columnwidth}@{}}
\toprule
\textbf{Abstract primitive} & \textbf{Realization} & \textbf{Reference} \\
\midrule
Tamper-resistant module \(\mathit{Aut}_c\); \(U\)-verified \(\mathsf{Sig}\), \(\mathsf{PRF}\) & WebAuthn with PRF extension; ES256/P-256 in this profile & \citep{webauthn2019,fido2prf2023} \\
Hash \(\mathsf{H}\) & SHA-256 & n/a \\
KDF (extract-and-expand, domain separation) & HKDF-SHA-256 & \citep{krawczyk2010hmac,chen2018kdfrec} \\
AEAD \(\mathsf{Enc}/\mathsf{Dec}\) & AES-GCM or (X)ChaCha20-Poly1305 & \citep{dworkin2007nist,nir2018chacha,arciszewski2020xchacha} \\
Key wrap \(\mathsf{Wrap}/\mathsf{Unwrap}\) & AES-KW / AES-KWP (or AEAD-as-wrap) & \citep{dworkin2012keywrap,schaad2002keywrap,housley2009kwp} \\
KEM \(\mathsf{Encap}/\mathsf{Decap}\) & HPKE (DHKEM-P256 / HKDF-SHA-256 / AEAD) & \citep{barnes2022hpke} \\
\bottomrule
\end{tabular}
\end{table}

\subsection{Cross-Device Channel via HPKE}
\label{sec:concrete-xdevice}

When \(U\) and \(T\) do not share a live TLS session (the agent runs on hosted infrastructure while \(U\) is on a phone, for example), Phase~II's channel assumptions (authenticity and confidentiality) must be realized by an alternative transport. \emph{This profile is transport-only}: it does not modify the abstract Phase~II grant (\S\ref{sec:phase2}); it only provides a confidential channel that delivers \(G\) intact from \(U\) to \(T\).

\emph{Authenticity of \(pk_T\) is required.} Confidentiality of \(W^\star\) (carried inside \(G\)) is security-critical: any party that observes \(W^\star\) can directly unwrap \([K]_{c^\star}\) from \(\Sigma\) and decrypt \(M\). Without an authentic \(pk_T\), HPKE Base admits an active man-in-the-middle: an adversary substituting its own \(pk_M\) for \(pk_T\) causes \(U\) to seal \(G\) to a key the adversary holds. \(U\) must therefore verify that \(pk_T\) belongs to the intended custodian. A concrete profile may obtain this authenticity through a signature on \(pk_T\) under a long-term \(T\)-instance key provisioned to \(U\)'s trusted client at registration; through binding \(pk_T\) to an authenticated session with the orchestration service that issued the cross-device link, which requires trusting that service for custodian-key binding; through an out-of-band channel integrity-protected against substitution, such as a QR code on a console authenticated to \(U\)~\citep{mccune2005seeing}; or through a mutually authenticated PAKE.

\emph{Construction.} Given an authentic \(pk_T\), \(U\) seals the abstract grant \(G = (o,\, r,\, \mathit{cid}_{c^\star},\, W^\star,\, \sigma^\star,\, \mathit{opt})\) to \(T\) using HPKE Base mode~\citep{barnes2022hpke}:
\[
  (\mathit{enc},\, \mathit{ct}_G) \gets \mathsf{HPKE.SealBase}\bigl(pk_T,\; \mathit{info},\; \varepsilon,\; G\bigr),
\]
with \(\mathit{info} = \mathsf{DS}_{\mathsf{xd}} \,\|\, r\) binding the freshness token into HPKE's key schedule (empty AAD). \(U\) transmits \((\mathit{enc},\, \mathit{ct}_G)\) to \(T\) as a CBOR envelope or equivalent, and \(T\) recovers \(G \gets \mathsf{HPKE.OpenBase}(sk_T,\, \mathit{enc},\, \mathit{info},\, \varepsilon,\, \mathit{ct}_G)\) before running Phase~II.3 redemption against \(\sigma^\star\) and \(\beta\) unchanged.

Under (A2, A5), HPKE Base provides IND-CCA2 sealed delivery to \(pk_T\); the \(\mathit{info}\)-bound \(r\) commits the channel to the freshness token, complementing the protocol-level single-use consumption of \(r\) at Phase~II.3. The channel does not supply authorization, which remains \(\sigma^\star\) inside \(G\): a relay that captures \((\mathit{enc},\, \mathit{ct}_G)\) cannot reuse it because \(r\) is consumed at \(T\) on first redemption (step~1), and substitution of \(o\) inside \(G\) fails \(\sigma^\star\) verification (step~3). Under the authenticity precondition on \(pk_T\), the profile does not require a live TLS tunnel between \(U\) and \(T\).

\subsection{Implementation and Evaluation}
\label{sec:implementation}

The \artifactnoun{} accompanying this paper implements the three protocol roles: a custodian library in Rust (about 4,500 lines) covering sealed-state management, grant validation, and consumption, and authorizer and requester libraries in TypeScript (about 1,600 lines) covering authenticator invocation and grant composition, exercised by conformance and protocol-flow test suites. The custodian realizes the profile of \S\ref{sec:concrete-primitives} with XChaCha20-Poly1305 sealing and HKDF-SHA-256 derivation. An end-to-end demonstration runs Phases I--III across separate custodian and requester processes.

\begin{table}[t]
\centering
\footnotesize
\caption{Per-operation latency over loopback HTTP (mean $\pm$ std, 100 iterations after warmup; software-emulated authenticator).}
\label{tab:bench}
\begin{tabular}{@{}l r@{}}
\toprule
\textbf{Operation} & \textbf{Latency [ms]} \\
\midrule
Direct bearer call (baseline) & $0.45 \pm 0.23$ \\
II.1 proposal and freshness issuance & $0.54 \pm 0.46$ \\
II.2 binding and signing (authorizer side) & $0.15 \pm 0.15$ \\
II.3+III.1 redemption and use & $0.52 \pm 0.19$ \\
End-to-end mediated call & $1.21 \pm 0.61$ \\
\bottomrule
\end{tabular}
\end{table}

\paragraph{Microbenchmark.}
Table~\ref{tab:bench} reports per-operation latency measured by the artifact's benchmark harness (Intel Xeon Platinum 8581C, loopback). A fully mediated call completes in 1.21\,ms and adds 0.76\,ms over a direct bearer-authenticated call on the same HTTP stack; under a standing grant (\S\ref{sec:opgrant}), the recurring per-call cost is bounded by the redemption-and-use leg (0.52\,ms). The grant request carries 509\,B more than the direct call. The authenticator is emulated in software, so real WebAuthn authenticator and user-interaction latency is excluded; custodian-side cryptography runs the reference library. Protocol overhead is thus negligible against network and ceremony latency.

\section{Security Analysis}
\label{sec:security}

We state \sudp's security as one main theorem over explicit assumptions and three interface preconditions (\S\ref{sec:security-ledger}), discharged by five propositions; Appendix~\ref{app:proofs} defines the security games, proves each proposition by reduction, and gives the axis-coverage table (Table~\ref{tab:axis-coverage}). Mechanized analysis (Tamarin~\citep{meier2013tamarin}, ProVerif~\citep{blanchet2001proverif}) is left to future work.

\subsection{Assumptions and Interface Preconditions}
\label{sec:security-ledger}

\paragraph{Assumptions.}
\begin{description}[leftmargin=2.5em,labelindent=0pt,topsep=2pt,itemsep=1pt]
  \item[(A1)] EUF-CMA security of the signature scheme \(\mathsf{Sig}\).
  \item[(A2)] Authenticated-encryption security of the AEAD scheme and key-wrap security of \((\mathsf{Wrap}, \mathsf{Unwrap})\).
  \item[(A3)] PRF security of the authenticator \(\mathit{Aut}_c\) (WebAuthn PRF extension) and of the \(\mathsf{KDF}\).
  \item[(A4)] Collision resistance of \(\mathsf{H}\) combined with deterministic, injective canonical serialization of \(o\).
  \item[(A5)] IND-CCA2 security of the KEM (HPKE).
  \item[(A6)] Authenticated and confidential channel \(U\!\to\!T\) for the Phase~I setup legs (credential enrollment and \(W_c\) transport) and for the Phase~II leg carrying \(W^\star\) (TLS~1.3, or the cross-device profile of \S\ref{sec:concrete-xdevice} under its authenticated-key precondition).
  \item[(A7)] Single-use freshness-state integrity at \(T\): \(r\) is consumed exactly once, and \(\mathsf{S}\), including the redeemed-grant record \(\rho\), is not adversarially modified between issuance and the end of the grant's validity window (no \(T\) memory compromise during this interval).
  \item[(A8)] (Conditional, used only by Proposition~\ref{prop:rfsauth}.) \(E\) supports rotation and revocation of authority-bearing secrets on the deployment's rotation cadence.
  \item[(A9)] Runtime hygiene at \(T\): fresh draws are CSPRNG-quality, and retired epoch keys (\(K\), \(W^\star\), peer wrapping keys) are erased after every wrap-epoch advance (\S\ref{sec:rotation}).
  \item[(A10)] (Conditional, used only by the in-state peer map of \S\ref{sec:rotation}.) Key-dependent-message security~\citep{black2002kdm,camenisch2001circular} of \((\mathsf{Enc}, \mathsf{Wrap})\) for that policy's key cycle (Appendix~\ref{app:csb}).
\end{description}

\paragraph{Interface preconditions.}
\begin{description}[leftmargin=2.5em,labelindent=0pt,topsep=2pt,itemsep=1pt]
  \item[(P1)] \emph{Canonical Operation Completeness.} Every authority-relevant field of the action exercised at \(E\) is represented in \(o.\mathsf{act}.\mathsf{scope}\), \(o.\mathsf{bind}\), or \(o.\mathsf{valid}\) (\S\ref{sec:opgrant}).
  \item[(P2)] \emph{Trusted Rendering Faithfulness.} \(\mathsf{Render}(o)\) at \(U\) is deterministic, complete with respect to the fields in \(o\), and produced inside \(U\)'s trusted client (\S\ref{sec:opgrant}).
  \item[(P3)] \emph{Execution Determinism.} The native call performed at \(E\) by \(T\) is a deterministic function of accepted \(o\), the protected material released inside \(T\), and any \(o\)-committed payload; \(R\) cannot supply authority-relevant fields after redemption (\S\ref{sec:opgrant}).
\end{description}

\subsection{Main Theorem}
\label{sec:security-main}

\begin{theorem}[\sudp{} solves the Agent Secret Use problem]\label{thm:asu}
Under (A1--A7, A9) and interface preconditions (P1--P3), \sudp{} enforces the four core security properties of \S\ref{sec:taxonomy}: AV, OB, UB, and CRC. By the criterion of \S\ref{sec:taxonomy}, \sudp{} therefore solves the Agent Secret Use problem (Definition~\ref{def:agentsecretuse}) against the baseline adversary \((\mathcal{A}^{R}, \mathcal{A}^{\mathrm{net}})\) of \S\ref{sec:adversary}. Concretely:
\begin{enumerate}[leftmargin=*,topsep=2pt,itemsep=1pt]
  \item Every Phase~III execution admitted by \(T\) is covered by a fresh, \(U\)-rooted, operation-bound, use-bounded grant (AV, OB, UB);
  \item For ASU-preserving operations (\S\ref{sec:phase3}), \(R\)'s view across all sessions consists of public proposal material, \(\mathsf{Release}(o)\), and recipient-sealed deliveries \(\pi\), and contains no artifact that confers reusable authority (CRC).
\end{enumerate}
\sudp{} additionally satisfies CSB under (A2, A3) and RFS-wrap under (A2, A3, A9), each plus (A10) under the in-state peer map; authority-level forward secrecy (RFS-auth) holds under the further assumption (A8). Runtime-memory confidentiality of \(T\) during the unwrap-use-zeroize window (CMC) is out of scope; it is the natural composition target for TEE-backed custodians.
\end{theorem}

\subsection{Supporting Propositions}
\label{sec:security-props}

\begin{proposition}[Authorization Integrity: AV $\wedge$ OB $\wedge$ UB]\label{prop:integrity}
Under (A1, A4, A6, A7) and (P1--P3), every Phase~III execution that \(T\) admits carries a signature \(\sigma^\star\) verifiable as a \(U\)-rooted authenticator assertion under \(\mathsf{Reg}[\mathit{cid}_{c^\star}]\) over \(\beta = \mathsf{H}(\mathsf{DS}_{\mathsf{bind}} \,\|\, r \,\|\, \mathsf{H}(o))\) (AV); no valid grant admits any \(o' \not\equiv_E o\) (OB); and every Phase~III execution stays within the bounds \(o.\mathsf{valid} = (\mathsf{expiry}, \mathsf{multiplicity})\) signed by \(\sigma^\star\) (UB).
\end{proposition}

\begin{proposition}[Secret Confidentiality: CRC $\wedge$ CSB]\label{prop:non-exposure}
In the trust model of \S\ref{sec:trust-model}: under (A2, A3, A6), for ASU-preserving operations (\S\ref{sec:phase3}), \(R\)'s view across all sessions is bounded by public proposal material, the authorized result form \(\mathsf{Release}(o)\), and sealed deliveries \(\pi\) (Proposition~\ref{prop:export}); it contains no value from which \(s\), \(K\), \(W_c\), or \(s_o\) can be recovered (CRC). Independently, under (A2, A3; A10 for the in-state peer map), an adversary obtaining the full persistent state \(\Sigma\) without a live \(U\)-verified \(\mathit{Aut}_c\) invocation cannot recover \(M\) (CSB).
\end{proposition}

\begin{proposition}[Recipient-Protected Export]\label{prop:export}
Given an accepted operation \(o\) satisfying Proposition~\ref{prop:integrity}, and under (A5, A2), for any ASU-preserving Phase~III.2 execution (a well-formed \(o\) whose \(\mathsf{bind}.\mathsf{recipient}\) identifies a principal outside \(R\)'s trust boundary), the delivery \(\pi\) reveals nothing about \(s_o\) to any party other than the holder of \(\mathit{sk}_{\mathsf{rcp}}\).
\end{proposition}

\begin{proposition}[Wrapping-Epoch Key Isolation]\label{prop:rfswrap}
Under (A2, A3, A9; A10 for the in-state peer map), a Phase~III.3 update yields wrapping-epoch key isolation: the post-rotation keys \((K', W^\star_{\mathsf{new}})\) are computationally independent of the pre-rotation keys \((K, W_{c^\star})\); compromise of one epoch's keys yields neither the other epoch's keys nor its retained ciphertexts.
\end{proposition}

\paragraph{Plaintext caveat.}
Wrapping-epoch key isolation does not by itself protect plaintext carried forward across rotations. The rewrap and pure-rotate sub-cases of Phase~III.3 explicitly preserve \(M' = M\) (\S\ref{sec:rotation}), so an adversary who decrypts either epoch's ciphertext recovers the same \(M\). Plaintext-level forward secrecy of authority-bearing material requires either (i) a Phase~III.3 update that retires the prior secret in \(M\), or (ii) composition with \(E\)-side authority rotation (Proposition~\ref{prop:rfsauth}).

\begin{proposition}[Authority-Level Forward Secrecy under E-side rotation]\label{prop:rfsauth}
Under (A8) and the assumptions of Proposition~\ref{prop:rfswrap}, compromise of one epoch's authority-bearing secret at \(E\) confers no authority over operations in any other epoch.
\end{proposition}

\subsection{Residual Limitations}
\label{sec:security-limits}

Five residual limitations (L1--L5) are explicit in the design.

\emph{(L1) Runtime compromise during consumption.} A fully compromised \(T\) runtime during legitimate consumption exfiltrates plaintext by construction: III.0 materializes \(K\) and \(M\) inside \(T\)'s memory, and no protocol-level mechanism prevents code executing inside \(T\) from reading them. Composition with confidential-computing runtimes is the natural remediation; zeroizing \(K\), \(W^\star\), and derived plaintext immediately after use narrows the window.

\emph{(L2) Authenticator compromise.} A fully compromised authenticator \(\mathit{Aut}_{c^\star}\) reduces Phase~II.2 to adversary-controlled authorization; authenticator security (A3) is an assumption, not a protocol-level property.

\emph{(L3) Dormant-credential forward-secrecy gap.} Lemma~\ref{cl:pcfs}'s forward secrecy is per-credential and activates only after a credential rotates. A credential that is registered but never participates in a state-update retains its initial salt indefinitely; historical PRF capture remains useful against its wrapping entry until the credential is exercised. This is intrinsic to any-of-\(N\) authorization without a global ceremony; deployments should drive scheduled rewraps for dormant credentials (\S\ref{sec:rotation}).

\emph{(L4) Peer-map cross-credential exposure.} Under the in-state peer-map construction, Phase~III.0 by any credential transiently exposes the current \(\{W_c\}_{c \in \mathsf{Creds}}\) to \(T\); a captured \(W_{c'}\) continues to unwrap subsequent rewraps until \(c'\) itself rotates. Deployments requiring stronger cross-credential isolation may substitute the all-present alternative of \S\ref{sec:rotation}.

\emph{(L5) Authorization is bounded by \(U\)'s judgment.} \sudp{} enforces the operations \(U\) signs; a dangerous but correctly rendered descriptor approved by \(U\) is executed. The semantic content of a conformant \(\mathsf{Release}(o)\) is similarly a deployment-policy concern outside protocol scope.

\section{Conclusion}
\label{sec:conclusion}

Agents that act autonomously on a user's behalf must be able to use the user's secrets without ever gaining reusable authority over them. We formalize this as the \emph{Agent Secret Use} problem and decompose it into seven security properties (AV, OB, UB for authorization integrity; CRC, CSB, CMC, RFS for secret confidentiality).

\sudp{} solves the problem by fusing authorization and access in a single authenticator ceremony: the invocation that signs the operation-bound grant is the same one that releases the key material unwrapping the secret, so no execution path reaches the secret without a reviewed operation. Six of the seven properties hold natively under standard cryptographic assumptions; the seventh, CMC, is closed by composition with a hardware-rooted runtime. Instantiated with commodity passkeys and standard primitives, the reference implementation adds sub-millisecond overhead per mediated call. In an LLM-based agent, the protocol mediates a call exactly when its execution would exercise user-enrolled authority-bearing material; the mediation is transparent to the agent.

Four directions remain open: formal verification under active adversaries; empirical evaluation of approval cost under realistic agent workloads; reconciling agent autonomy with grant granularity, by data-driven grant bounds or composition with policy systems; and authenticator primitives for non-human principals, allowing chains whose authorizer is itself an agent under further-rooted authority.

\medskip\noindent\textbf{Generative AI disclosure.} Generative AI tools assisted in drafting and editing the prose; all technical content and claims originate with the authors, who take full responsibility for the final text.

\section*{Ethics Considerations}
This work specifies and analyzes a defensive authorization protocol; it involves no human-subjects research and no collection of personal data. The weaknesses discussed are structural properties of mechanism classes, stated from public specifications and documentation; the paper discloses no vulnerability in an identifiable deployment, so coordinated disclosure was not applicable.

\bibliographystyle{IEEEtranN}
\bibliography{references}

\appendices
\section{Design Rationale}
\label{sec:design-choices}

Several \sudp{} design decisions are consequential but otherwise scattered across the construction. This section consolidates the architectural-level choices to make their rationale visible together; smaller decisions (domain-separation strings, concrete primitive selections) are discussed inline at the point of construction.

\paragraph{Per-grant \(U\)-signed authorization committed to \(\mathsf{H}(o)\).}
The redeemed grant \(\rho\) is cryptographically bound to a specific operation via \(\mathsf{H}(o)\) at the authenticator, not to a stored consent reused across token lifetimes. This lifts anti-replay and anti-substitution from engineering discipline (TTLs, revocation lists, nonce tracking in each tool) to protocol-level guarantees: any captured artifact, within any time window, is redeemable only against the operation it was signed for. \sudp{} carries the execution budget in \(o.\mathsf{valid}.\mathsf{multiplicity} \in \mathbb{Z}^{+} \cup \{\infty\}\), which $U$ signs as part of the grant. \emph{Single-use} grants (\(\mathsf{multiplicity}{=}1\)) match transaction-confirmation ceremonies (PSD2 dynamic linking, banking 2FA): consumed on first use, suited to high-authority actions whose misuse already warrants per-action human oversight. \emph{TTL-bounded standing} grants (\(\mathsf{multiplicity}{=}\infty\), bounded by \(\mathsf{expiry}\)) match a multi-entry visa with a fixed validity window: $U$ explicitly opts into repeated use within a $U$-declared rule, suited to low-risk repeat patterns (\eg, inbox reads) where per-action confirmation is operationally untenable. In both modes, the bounds are committed by \(U\)'s per-grant signature, not delegated to an authorization server's policy.

\paragraph{Authenticator-bound PRF for wrapping material.}
Wrapping keys are derived from authenticator-bound PRF outputs (\(y_c = \mathsf{PRF}_c(\eta_c)\) inside \(\mathit{Aut}_c\)) rather than from passwords, server-held master keys, or envelope-encryption KMS services. The unlock material therefore lives exclusively in \(U\)'s authenticator hardware: neither the server's persistent state nor its runtime memory can reconstruct a wrapping key without a live \(U\)-verified invocation. This is what distinguishes \sudp's satisfaction of CSB from schemes whose ``encrypted at rest'' claim collapses when an attacker captures the server together with its unlock material (\S\ref{sec:positioning}). The cost is a WebAuthn PRF-capable authenticator, or an equivalent authenticator-bound PRF, as a deployment prerequisite. Ordinary passkey availability is broad, but PRF support remains a profile requirement that deployments must check.

\paragraph{Rewrap policy at rotation.}
A \sudp{} profile inherits the recoverability semantics of its authenticator family: if all enrolled credentials are lost, $M$ becomes unrecoverable through \sudp{}. Deployments that need recovery must provision it as an additional enrolled credential (a recovery passkey, a second hardware authenticator, or an enterprise-controlled escrow credential whose use is itself an \sudp{}-authorized lifecycle operation), and never as a backend master-key bypass, which would surrender CSB. Each enrolled credential therefore holds its own wrapping key \(W_c\), and every Phase~III.3 state-update by one credential must rewrap the new \(K'\) under all peers (\S\ref{sec:rotation}). Three rewrap policies meet this requirement, summarized in Table~\ref{tab:recov}. \emph{In-state peer map} (default) keeps each rewrap single-invocation, at the cost of a transient \(\{W_c\}\) exposure to \(T\) during III.0, a per-peer forward-secrecy gap until that peer itself rotates, and the key-cycle assumption (A10) of \S\ref{sec:security-ledger}. \emph{All-present rotation} removes the exposure but forces every credential online at each rewrap, which is ill-suited to continuous agent workflows. \emph{Append-only chain} supports offline catch-up at unbounded chain storage but is not forward-secret: a captured \(K_j\) walks the chain forward. The default favors single-invocation rewrap for agent workflows that rotate on every write; deployments needing stronger cross-credential isolation may substitute all-present, or adopt append-only as an offline-catchup auxiliary with the FS caveat above.

\begin{table}[h]
\centering
\footnotesize
\setlength{\tabcolsep}{3pt}
\caption{Rewrap policies for multi-credential \sudp{} rotation.}
\label{tab:recov}
\resizebox{\columnwidth}{!}{%
\begin{tabular}{@{}l l l l l@{}}
\toprule
\textbf{Policy} & \textbf{Online invocations} & \textbf{Extra storage} & \textbf{Peer-key exposure} & \textbf{FS profile} \\
\midrule
In-state peer map (default) & 1 (single-invocation) & $O(|\mathsf{Creds}|)$ in $M$ & transient during III.0 & Lemma~\ref{cl:pcfs} per credential, after $c$ rotates \\
All-present rotation & all credentials & none & none & Lemma~\ref{cl:pcfs} per credential, independently \\
Append-only chain & 1 (single-invocation) & unbounded (chain grows) & none & \emph{not} FS: old $K_j$ walks forward \\
\bottomrule
\end{tabular}%
}
\end{table}

\paragraph{Minimality of the binding hash.}
The binding hash commits only to freshness and the operation:
\[
  \beta = \mathsf{H}(\mathsf{DS}_{\mathsf{bind}} \,\|\, r \,\|\, \mathsf{H}(o)).
\]
The credential identifier is intentionally omitted: \(\mathit{cid}_{c^\star}\) is delivered in \(G\) alongside \(\sigma^\star\), and verification through \(\mathsf{Reg}[\mathit{cid}_{c^\star}] \to \mathit{pk}_{c^\star}\) already enforces credential binding under EUF-CMA; restating it inside \(\beta\) would be redundant. The custodian's identity and the server's TLS certificate hash are likewise omitted: those are bound by the authenticated channel carrying the grant rather than by the application-layer authorization challenge. Re-binding fields already bound elsewhere trades clarity for bloat, and \(\beta\) follows standard key-exchange transcript discipline of carrying only what the surrounding structure does not already commit~\citep{krawczyk2005hmqv}.

\paragraph{Per-write rotation by design.}
Wrapping-epoch rotation is mandated by the protocol rather than left to operator discipline: every Phase~III.3 state-update samples a fresh \(K'\) and a fresh next-salt \(\eta^{\mathsf{next}}_{c^\star}\) (\S\ref{sec:rotation}). Coupling rotation to every write removes the need for a separate rotation ceremony and delivers wrapping-epoch forward secrecy unconditionally under key-erasure assumptions (Proposition~\ref{prop:rfswrap}, Lemma~\ref{cl:pcfs}). Authority-level rotation (rotating the underlying secret at \(E\)) is kept separate, runs on a deployment-defined cadence, and depends on \(E\)-side revocation support (Proposition~\ref{prop:rfsauth}); the separation lets wrap-side FS hold independently of \(E\)'s rotation API. The per-write cost is one CSPRNG draw, one key derivation, and \(|\mathsf{Creds}|\) rewraps, which is marginal for agent workflows whose write volumes are typically low per user.

\paragraph{Extensibility of the dispatch vocabulary.}
The vocabulary \(\{\mathsf{use}, \mathsf{export}, \mathsf{lifecycle}\}\) covers the canonical consumption modes; concrete profiles MAY define additional dispatch types provided each new type (i)~preserves the channel binding \(\beta = \mathsf{H}(\mathsf{DS}_{\mathsf{bind}}\,\|\,r\,\|\,\mathsf{H}(o))\) and the grant-validation steps of II.3 unchanged, (ii)~inherits or explicitly restates its AV/OB/UB stance, and (iii)~specifies its non-disclosure stance against \(R\), stating what plaintext-bearing material, if any, leaves \(T\). Profile-defined dispatch types remain within the abstract protocol when they meet these conditions and do not require modification of Phases~I or~II. Extensions that weaken non-disclosure relative to the canonical modes are admissible at the profile level only as deliberate trade-offs and must be surfaced as such, both at authorization time to \(U\) and in the profile's stated security claims.

\paragraph{Submission topologies.}
\label{sec:concrete-topologies}
The sealed grant may reach \(T\) directly, over TLS~1.3 or the envelope above, or through a relay that forwards the ciphertext. The natural relay in an agentic deployment is \(R\) itself, which lets the agent retry, batch, or schedule submission while observing nothing beyond the HPKE ciphertext sealed to \(pk_T\). Both topologies carry the same guarantees, since \(\sigma^\star\) binds the operation independently of the carrier and \(r\) is single-use. Direct submission keeps the session under \(U\)'s control but requires \(T\) to be reachable at signing time; relayed submission decouples signing from delivery at the cost of routing through \(R\)'s trust boundary.

\section{Formal Security Model and Proofs}
\label{app:proofs}

This appendix restates the guarantees of \S\ref{sec:security} as adversarial games and proves the supporting propositions by reduction to the assumptions of \S\ref{sec:security-ledger}, in the game-based tradition of entity authentication~\citep{bellare1993eakd} and the game-sequence proof style~\citep{shoup2004games,bellare2006code}. Appendix~\ref{app:model} fixes the execution model shared by all games. Theorem~\ref{thm:auth} proves Proposition~\ref{prop:integrity}; Lemma~\ref{lem:view} and Theorem~\ref{thm:exphide} prove the CRC clause of Proposition~\ref{prop:non-exposure} and Proposition~\ref{prop:export}; Theorem~\ref{thm:csb} proves the CSB clause; Theorem~\ref{thm:rfs} proves Proposition~\ref{prop:rfswrap} by giving Lemma~\ref{cl:pcfs} its game form. Table~\ref{tab:axis-coverage} maps the taxonomy axes to these results.

Throughout, \(\lambda\) is the security parameter, all algorithms run in probabilistic polynomial time (PPT) in \(\lambda\), and freshness tokens are \(\lambda\)-bit strings. Each assumption of \S\ref{sec:security-ledger} contributes an advantage term: \(\epsilon_{\mathsf{sig}}\) for the EUF-CMA experiment of \(\mathsf{Sig}\) (A1), \(\epsilon_{\mathsf{ae}}\) and \(\epsilon_{\mathsf{wrap}}\) for the AEAD and key-wrap experiments (A2), \(\epsilon_{\mathsf{prf}}\) and \(\epsilon_{\mathsf{kdf}}\) for the PRF experiments of \(\mathit{Aut}_c\) and \(\mathsf{KDF}\) (A3), \(\epsilon_{\mathsf{cr}}\) for collision resistance of \(\mathsf{H}\) (A4), \(\epsilon_{\mathsf{kem}}\) for the IND-CCA2 experiment of the KEM (A5), and \(\epsilon_{\mathsf{kdm}}\) for the key-dependent-message experiment of (A10), stated in Appendix~\ref{app:csb}. Each \(\epsilon\) denotes the maximum advantage over adversaries with resources comparable to the reductions constructed below. The model covers one custodian and one protected state; batch grants substitute \(\mathsf{H}(\mathsf{ops})\) for \(\mathsf{H}(o)\) throughout (\S\ref{sec:phase2}) and change no argument.

\subsection{Execution Model}
\label{app:model}

A challenger runs the authorizer \(U\), its authenticator modules \(\mathit{Aut}_c\), and the custodian \(T\) honestly, following Phases I--III. At enrollment, each \(\mathit{Aut}_c\) samples its signing key and PRF key independently (Phase~I.1); the reductions below use this to simulate one half of the module while embedding a challenge in the other. The adversary \(\mathcal{A}\) plays the baseline \((\mathcal{A}^{R}, \mathcal{A}^{\mathrm{net}})\) of \S\ref{sec:adversary}: it controls \(R\) and every network leg except the \(U\!\to\!T\) grant leg, which A6 models as authenticated and confidential, so the challenger delivers grants internally and \(\mathcal{A}\) observes only their public part. The Phase~I setup legs of A6 are likewise challenger-internal. \(T\)'s code and bookkeeping are honest, per the restricted custodian trust of \S\ref{sec:roles}; A7 states the same boundary in deployment terms, and A9 supplies the uniformity of fresh draws. The interface preconditions enter as modeling conventions: \(o\) is the canonical tuple of Definition~\ref{def:op} and carries every authority-relevant field (P1); the rendering \(U\) approves is the canonical \(o\) itself (P2); execution at \(E\) is an abstract deterministic map \(\mathsf{Rel}\), a parameter of the game, with \(\mathsf{Release}(o) = \mathsf{Rel}(o, s_o)\) (P3). Time is a counter \(\mathcal{A}\) advances; \(T\) reads it when enforcing \(o.\mathsf{valid}\).

The challenger keeps an authorization log \(\mathcal{L}_{\mathsf{auth}}\) of triples \((o, r, c)\) signed by \(U\) and an acceptance log \(\mathcal{L}_{\mathsf{acc}}\) of triples accepted at II.3. \(\mathcal{A}\) may invoke, in any order and polynomially often:
\begin{description}[leftmargin=1em,topsep=2pt,itemsep=1pt]
  \item[\textsc{enroll}()] enrolls a fresh credential \(c\) (Phase I.1--I.2) and returns the public \((\mathit{cid}_c, \mathit{pk}_c, \eta_c)\).
  \item[\textsc{req}(\(o\))] runs Phase II.1 with \(\mathcal{A}\) as \(R\): \(T\) draws \(r \stackrel{\$}{\gets} \{0,1\}^\lambda\), stores it in \(\mathsf{S}\), and returns the conveyance \((o, r, \{(\mathit{cid}_c, \eta_c)\})\).
  \item[\textsc{grant}(\(o, r, c\))] runs Phase II.2 on an adversary-chosen conveyance, modeling tampering on the \(T \to U\) leg, with acting credential \(c\): \(U\) approves, computes \(\beta = \mathsf{H}(\mathsf{DS}_{\mathsf{bind}} \,\|\, r \,\|\, \mathsf{H}(o))\), obtains \(\sigma^\star\), \(W^\star\), and (for rotation-class \(o\)) \(\mathit{opt}\) from \(\mathit{Aut}_c\), and sends the grant toward \(T\) on the A6 channel. The challenger appends \((o, r, c)\) to \(\mathcal{L}_{\mathsf{auth}}\) and returns the public grant part \((o, r, \mathit{cid}_c, \sigma^\star)\); \(W^\star\) and \(\mathit{opt}\) are withheld. \(U\) approves every query: authorization policy lies outside the model (L5), and a restrictive \(U\) only shrinks \(\mathcal{A}\)'s options.
  \item[\textsc{deliver}(\(i\))] delivers the \(i\)-th grant unmodified to \(T\), which runs II.3; on acceptance the challenger appends the grant's \((o, r, c)\) to \(\mathcal{L}_{\mathsf{acc}}\). The oracle returns \(T\)'s decision; \(\mathcal{A}\) chooses delivery order, delay, and omission.
  \item[\textsc{redeem}(\(\tilde{G}\))] submits a grant \(\tilde{G} = (\tilde{o}, \tilde{r}, \mathit{cid}_{\tilde{c}}, \tilde{W}, \tilde{\sigma}, \tilde{\mathit{opt}})\) of \(\mathcal{A}\)'s own construction to \(T\), with the same acceptance bookkeeping; this models injection toward \(T\) from any network position.
  \item[\textsc{exec}(\(j\))] runs one Phase III execution under the \(j\)-th accepted grant: \(T\) executes III.0 and the arm selected by \(o_j.\mathsf{act}.\mathsf{type}\), enforcing \(o_j.\mathsf{valid}\) and aborting on \(\mathsf{Unwrap}\) or \(\mathsf{Dec}\) failure. The oracle returns \(\mathsf{Release}(o_j)\) for \(\mathsf{use}\), the sealed delivery \(\pi_j\) for \(\mathsf{export}\), and an acknowledgment for lifecycle operations.
\end{description}

\subsection{Authorization Soundness: AV, OB, UB}
\label{app:auth}

\begin{definition}[Authorization-soundness game]\label{def:gauth}
\(\mathbf{Exp}^{\mathsf{auth}}(\mathcal{A})\) runs the model of Appendix~\ref{app:model}. \(\mathcal{A}\) wins if at any point (i)~\(T\) accepts a grant whose triple \((\tilde{o}, \tilde{r}, \tilde{c})\) is absent from \(\mathcal{L}_{\mathsf{auth}}\), (ii)~\(T\) accepts two grants carrying the same token value \(\tilde{r}\), or (iii)~a Phase III execution violates the accepted grant's \(o.\mathsf{valid}\) bounds. The advantage \(\mathbf{Adv}^{\mathsf{auth}}(\mathcal{A})\) is \(\Pr[\mathcal{A} \text{ wins}]\).
\end{definition}

Condition (i) is stated at acceptance rather than execution: every Phase III execution runs under an accepted grant (III.0), so acceptance-level soundness is the stronger statement. Condition (iii) has probability zero, because the challenger's \(T\) enforces \(o.\mathsf{valid}\) by bookkeeping; the game's cryptographic content is (i) and (ii).

\begin{theorem}[Authorization soundness]\label{thm:auth}
For every PPT \(\mathcal{A}\) making at most \(q_g\) \textsc{req} queries against a model with at most \(N\) enrolled credentials,
\[
  \mathbf{Adv}^{\mathsf{auth}}(\mathcal{A}) \;\le\; N\,\epsilon_{\mathsf{sig}} \,+\, \epsilon_{\mathsf{cr}} \,+\, q_g^2/2^{\lambda}.
\]
\end{theorem}
\begin{proof}
We bound each win condition; the bounds sum by a union bound.

\emph{Condition (ii).} Each acceptance consumes its token from \(\mathsf{S}\), so two acceptances of the same value \(\tilde{r}\) require \(T\) to have drawn \(\tilde{r}\) twice among at most \(q_g\) draws. Under A9 the draws are uniform \(\lambda\)-bit strings, so this event has probability at most \(q_g^2/2^{\lambda}\).

\emph{Condition (i).} Suppose \(T\) accepts \((\tilde{o}, \tilde{r}, \tilde{c}) \notin \mathcal{L}_{\mathsf{auth}}\). Acceptance includes \(\mathsf{Vrfy}_{\mathsf{Reg}[\mathit{cid}_{\tilde{c}}]}(\tilde{\beta}, \tilde{\sigma}) = 1\) for \(\tilde{\beta} = \mathsf{H}(\mathsf{DS}_{\mathsf{bind}} \,\|\, \tilde{r} \,\|\, \mathsf{H}(\tilde{o}))\). Let \(B_{\tilde{c}}\) denote the set of challenges \(\mathit{Aut}_{\tilde{c}}\) signed during \textsc{grant} queries.

\emph{Case \(\tilde{\beta} \notin B_{\tilde{c}}\).} The reduction \(\mathcal{B}_1\) samples an index uniformly from the at most \(N\) credentials, embeds its EUF-CMA challenge key as that credential's \(\mathit{pk}\), and simulates the module's two halves independently: signing via its signing oracle, and the PRF via a key \(\mathcal{B}_1\) samples itself, which is sound because the module's two keys are independent draws (Appendix~\ref{app:model}). All other credentials run honestly. If the win occurs at the guessed credential, \((\tilde{\beta}, \tilde{\sigma})\) verifies under the challenge key and \(\tilde{\beta}\) was never queried to the signing oracle, so \(\mathcal{B}_1\) outputs it as a forgery. This case occurs with probability at most \(N \epsilon_{\mathsf{sig}}\).

\emph{Case \(\tilde{\beta} \in B_{\tilde{c}}\).} Some \((o, r, \tilde{c}) \in \mathcal{L}_{\mathsf{auth}}\) satisfies \(\mathsf{H}(\mathsf{DS}_{\mathsf{bind}} \,\|\, r \,\|\, \mathsf{H}(o)) = \tilde{\beta}\) with \((o, r) \neq (\tilde{o}, \tilde{r})\). If \((r, \mathsf{H}(o)) \neq (\tilde{r}, \mathsf{H}(\tilde{o}))\), the two outer inputs are distinct and the reduction \(\mathcal{B}_2\) outputs them as an \(\mathsf{H}\)-collision. Otherwise \(r = \tilde{r}\) and \(\mathsf{H}(o) = \mathsf{H}(\tilde{o})\) with \(o \neq \tilde{o}\); canonical serialization is injective (A4), so the serializations of \(o\) and \(\tilde{o}\) are distinct and form an inner \(\mathsf{H}\)-collision. This case occurs with probability at most \(\epsilon_{\mathsf{cr}}\).
\end{proof}

\begin{proof}[Proof of Proposition~\ref{prop:integrity}]
By Theorem~\ref{thm:auth}, except with the stated advantage, every acceptance at \(T\) carries a triple of \(\mathcal{L}_{\mathsf{auth}}\). \emph{AV}: the matching log entry is an authorization event produced by \(U\)'s authenticator over \(\beta\), and \(\mathsf{Reg}\) binds \(\mathit{cid}_{c^\star}\) to a key enrolled over the authenticated Phase~I channel (A6), so the event is attributable to \(U\) and re-verifiable from the public transcript. \emph{OB}: the accepted \(\tilde{o}\) equals the \(o\) of its matching log entry, which by P2 is the operation \(U\) rendered and approved; by P3, no authority-relevant field enters between authorization and execution. \emph{UB}: each token supports one acceptance (condition ii), and \(T\) enforces expiry and multiplicity at every execution (condition iii), so execution stays within the \(o.\mathsf{valid}\) that \(\sigma^\star\) committed to.
\end{proof}

\subsection{Requester View and Export Hiding: CRC}
\label{app:crc}

\begin{lemma}[View determination]\label{lem:view}
In the model of Appendix~\ref{app:model}, \(\mathcal{A}\)'s view consists of its own coins and the sequence
\(\mathcal{V} := \langle\)enrollment outputs \((\mathit{cid}_c, \mathit{pk}_c, \eta_c)\), conveyances \((o, r, \{(\mathit{cid}_c, \eta_c)\})\), public grant parts \((o, r, \mathit{cid}_c, \sigma^\star)\), acceptance decisions, \(\mathsf{Release}\) outputs, sealed deliveries \(\pi\), lifecycle acknowledgments\(\rangle\). In particular, no oracle returns \(W^\star\), \(\mathit{opt}\), \(K\), any \(W_c\), or any function of \(M\) other than \(\mathsf{Release}(o)\) and \(\pi\).
\end{lemma}
\begin{proof}
By inspection of the six oracle interfaces of Appendix~\ref{app:model}: each return value is listed there, the A6 channel withholds \(W^\star\) and \(\mathit{opt}\) from the public grant part, \(\Sigma\) is challenger-internal, and III.0's intermediate values \(K\), \(M\), and \(s_o\) are consumed inside \(T\)'s arm of \textsc{exec}.
\end{proof}

Lemma~\ref{lem:view} reduces CRC to the two components of \(\mathcal{V}\) that depend on protected material: \(\mathsf{Release}\) outputs, whose content is the authorized result form and whose semantics are \(U\)'s decision (L5), and sealed deliveries, which Theorem~\ref{thm:exphide} shows carry no information about their payload. The remaining components are public protocol material: in particular \(\sigma^\star\) binds only the public \(\beta\), is produced by the signing key, which is independent of the PRF key inside \(\mathit{Aut}_{c^\star}\) (Appendix~\ref{app:model}), and its replay is excluded by Theorem~\ref{thm:auth}.

\begin{definition}[Export-hiding game]\label{def:gexp}
\(\mathbf{Exp}^{\mathsf{exp}\text{-}b}(\mathcal{A})\) extends the model of Appendix~\ref{app:model} with a challenger-generated recipient key pair; \(\mathcal{A}\) receives \(\mathit{pk}_{\mathsf{rcp}}\) and may name it in \(o.\mathsf{bind}.\mathsf{recipient}\). For \(b = 0\), \textsc{exec} on an export with recipient \(\mathit{pk}_{\mathsf{rcp}}\) returns \(\pi = \mathsf{KEMSeal}(\mathit{pk}_{\mathsf{rcp}}, s_o)\) as in Phase III.2; for \(b = 1\), it returns \(\mathsf{KEMSeal}(\mathit{pk}_{\mathsf{rcp}}, 0^{|s_o|})\). \(\mathcal{A}\) outputs a bit, and its advantage is \(|\Pr[1 \gets \mathbf{Exp}^{\mathsf{exp}\text{-}0}(\mathcal{A})] - \Pr[1 \gets \mathbf{Exp}^{\mathsf{exp}\text{-}1}(\mathcal{A})]|\).
\end{definition}

\begin{theorem}[Export hiding]\label{thm:exphide}
For every PPT \(\mathcal{A}\) triggering at most \(q_x\) export executions to \(\mathit{pk}_{\mathsf{rcp}}\), the advantage in Definition~\ref{def:gexp} is at most \(q_x(\epsilon_{\mathsf{kem}} + \epsilon_{\mathsf{kdf}} + \epsilon_{\mathsf{ae}})\).
\end{theorem}
\begin{proof}
A hybrid over the \(q_x\) exports; each hop replaces one real sealing with a sealing of \(0^{|s_o|}\) through three game steps. First, replace the encapsulated key \(K_{\mathsf{d}}\) by an independent uniform key: \(\mathit{sk}_{\mathsf{rcp}}\) is never used in the game, so the reduction needs no decapsulation oracle and the CPA fragment of A5 suffices, at cost \(\epsilon_{\mathsf{kem}}\); A5's full IND-CCA2 strength covers deployments in which \(\mathit{sk}_{\mathsf{rcp}}\) also answers decapsulation queries outside the game. Second, replace \(k_{\mathsf{d}} = \mathsf{KDF}(K_{\mathsf{d}}; \bot, \mathsf{H}(o))\) by a uniform key, at cost \(\epsilon_{\mathsf{kdf}}\). Third, with \(k_{\mathsf{d}}\) uniform and used only as the AEAD key of \(\delta\), swap the plaintext \(s_o\) for \(0^{|s_o|}\) under the same associated data \(\mathsf{H}(o)\), at cost \(\epsilon_{\mathsf{ae}}\). Summing over the \(q_x\) hops gives the bound.
\end{proof}

\begin{proof}[Proof of Proposition~\ref{prop:non-exposure}, CRC clause]
By Lemma~\ref{lem:view}, \(\mathcal{A}\)'s view across all sessions is the public sequence \(\mathcal{V}\); by Theorem~\ref{thm:exphide}, replacing every sealed delivery for a recipient outside \(R\)'s trust boundary with a sealing of a constant string changes the view distribution by at most \(q_x(\epsilon_{\mathsf{kem}} + \epsilon_{\mathsf{kdf}} + \epsilon_{\mathsf{ae}})\). The view therefore reveals nothing about \(s\), \(K\), \(W_c\), or \(s_o\) beyond what the authorized \(\mathsf{Release}\) outputs disclose, which is the CRC clause for ASU-preserving operations.
\end{proof}

\begin{proof}[Proof of Proposition~\ref{prop:export}]
Secrecy of \(\pi\) against every party except the holder of \(\mathit{sk}_{\mathsf{rcp}}\) is Theorem~\ref{thm:exphide}. Recipient substitution is excluded by Theorem~\ref{thm:auth}: \(\mathit{pk}_{\mathsf{rcp}}\) lies in \(o.\mathsf{bind}\), so an export toward a different recipient is a different \(o\) and requires its own authorization event. Splicing a captured \(\pi\) into another operation's transcript fails because both the delivery-key derivation and the associated data of \(\delta\) bind \(\mathsf{H}(o)\) (Phase III.2). A grant naming \(R\) itself as recipient is a deliberate ownership transfer outside ASU non-disclosure (\S\ref{sec:phase3}).
\end{proof}

\subsection{Storage Breach: CSB}
\label{app:csb}

\begin{definition}[Storage-breach game]\label{def:gcsb}
In \(\mathbf{Exp}^{\mathsf{csb}\text{-}b}(\mathcal{A})\), the adversary outputs two protected states \((M^0, M^1)\) of equal shape (the same entry set, with entries of equal lengths) and a credential count \(N\); the challenger runs Phase~I over \(M^b\) with \(N\) credentials and hands \(\mathcal{A}\) the resulting \(\Sigma = (C, \{(\mathit{cid}_c, \eta_c, [K]_c)\}, \mathsf{Reg}, \mathit{ver})\); \(\mathcal{A}\) outputs a bit. The advantage is \(|\Pr[1 \gets \mathbf{Exp}^{\mathsf{csb}\text{-}1}(\mathcal{A})] - \Pr[1 \gets \mathbf{Exp}^{\mathsf{csb}\text{-}0}(\mathcal{A})]|\).
\end{definition}

The game is offline: it grants no protocol oracles after the breach, matching CSB's condition that no live \(U\)-verified invocation occurs (\S\ref{sec:taxonomy}). Epoch interleavings, where key material from one epoch meets state from another, are the subject of Appendix~\ref{app:rfs}. Under the all-present rewrap policy (\S\ref{sec:rotation}, \S\ref{sec:design-choices}), \(M\) carries no in-state wrapping material and Theorem~\ref{thm:csb} applies as stated. The default in-state peer map stores \(\mathsf{Peer} \subseteq M\) (\S\ref{sec:rotation}), closing a key cycle between \(\mathsf{Enc}_K(M)\) and \(\mathsf{Wrap}_{W_c}(K)\) that A2 does not cover; A10 asserts that \((\mathsf{Enc}, \mathsf{Wrap})\) remains secure for exactly this cycle. The A10 experiment: for adversary-chosen \(N\), uniform \(K\) and \(\{W_c\}_{c \in [N]}\), and adversary-chosen equal-shape \((M^0, M^1)\) whose peer-map entries the challenger overwrites with the \(W_c\) identically in both states, the tuple \(\bigl(\{\mathsf{Wrap}_{W_c}(K)\}_{c}, \mathsf{Enc}_K(M^b; \mathit{ver})\bigr)\) hides \(b\) with advantage at most \(\epsilon_{\mathsf{kdm}}\). Under A10, Theorem~\ref{thm:csb} extends to the in-state peer map: the first two hybrid steps of its proof make every \(W_c\) uniform without touching the cycle, after which \(\mathcal{A}\)'s remaining \(M^b\)-dependent input is the A10 tuple, so the wrap and encryption steps collapse into a single A10 reduction and the bound becomes \(2N(\epsilon_{\mathsf{prf}} + \epsilon_{\mathsf{kdf}}) + \epsilon_{\mathsf{kdm}}\); the same substitution extends Theorem~\ref{thm:rfs} with bound \(2(\epsilon_{\mathsf{prf}} + \epsilon_{\mathsf{kdf}}) + \epsilon_{\mathsf{kdm}} + 2^{1-\lambda}\). Cycles of this shape are routine in deployed designs such as full-disk encryption and password managers, with no known attack on practical AEAD schemes; the all-present policy avoids A10 entirely.

\begin{theorem}[Storage-breach indistinguishability]\label{thm:csb}
For every PPT \(\mathcal{A}\), the advantage in Definition~\ref{def:gcsb} is at most \(2N(\epsilon_{\mathsf{prf}} + \epsilon_{\mathsf{kdf}} + \epsilon_{\mathsf{wrap}}) + \epsilon_{\mathsf{ae}}\).
\end{theorem}
\begin{proof}
A sequence of games from \(\mathbf{Exp}^{\mathsf{csb}\text{-}0}\) to \(\mathbf{Exp}^{\mathsf{csb}\text{-}1}\). The public components \(\mathsf{Reg}\), \(\{\mathit{cid}_c\}\), \(\{\eta_c\}\), and \(\mathit{ver}\) are generated independently of \(M^b\) and are identically distributed throughout. Starting from the real game over \(M^0\): first, for each credential in turn, replace \(y_c = \mathsf{PRF}_c(\eta_c)\) by a uniform value; each authenticator's PRF key is distinct and evaluated once here, so the \(N\) hops cost \(N\epsilon_{\mathsf{prf}}\). Second, replace each \(W_c = \mathsf{KDF}(y_c; \eta_c, \cdot)\) by a uniform key, at cost \(N\epsilon_{\mathsf{kdf}}\). Third, with each \(W_c\) uniform and used only to produce \([K]_c\), replace each \([K]_c\) by the wrap of an independent key, at cost \(N\epsilon_{\mathsf{wrap}}\); after this step \(K\) appears in \(\mathcal{A}\)'s input only as the key of \(C\). Fourth, replace \(C = \mathsf{Enc}_K(M^0; \mathit{ver})\) by \(\mathsf{Enc}_K(M^1; \mathit{ver})\), at cost \(\epsilon_{\mathsf{ae}}\). Walking the first three steps back with \(M^1\) in place restores the real game over \(M^1\) and costs the same \(N(\epsilon_{\mathsf{prf}} + \epsilon_{\mathsf{kdf}} + \epsilon_{\mathsf{wrap}})\) again, giving the stated bound.
\end{proof}

\begin{proof}[Proof of Proposition~\ref{prop:non-exposure}, CSB clause]
An adversary that recovers \(M\) from \(\Sigma\) distinguishes \(\mathbf{Exp}^{\mathsf{csb}\text{-}0}\) from \(\mathbf{Exp}^{\mathsf{csb}\text{-}1}\) by comparing the recovered value against its own \((M^0, M^1)\), so Theorem~\ref{thm:csb} bounds its success probability.
\end{proof}

\subsection{Rotation Forward Secrecy: RFS}
\label{app:rfs}

\begin{table}[!t]
\centering
\footnotesize
\setlength{\tabcolsep}{4pt}
\caption{Taxonomy-axis coverage.}
\label{tab:axis-coverage}
\begin{tabular}{@{}l l l@{}}
\toprule
\textbf{Axis} & \textbf{Established by} & \textbf{Assumptions} \\
\midrule
AV & Prop.~\ref{prop:integrity}, Thm.~\ref{thm:auth} & A1, A4, A6, P1--P2 \\
OB & Prop.~\ref{prop:integrity}, Thm.~\ref{thm:auth} & A1, A4, P1--P3 \\
UB & Prop.~\ref{prop:integrity}, Thm.~\ref{thm:auth} & A1, A4, A7, A9 \\
CRC & Props.~\ref{prop:non-exposure}, \ref{prop:export}; Thm.~\ref{thm:exphide} & A2, A3, A5, A6 \\
CSB & Prop.~\ref{prop:non-exposure}, Thm.~\ref{thm:csb} & A2, A3 (+A10) \\
RFS & Props.~\ref{prop:rfswrap}, \ref{prop:rfsauth}; Thm.~\ref{thm:rfs} & A2, A3, A9 (+A8, +A10) \\
\midrule
CMC & out of scope (L1) & n/a \\
\bottomrule
\end{tabular}

\smallskip
\parbox{\columnwidth}{\scriptsize Each axis (\S\ref{sec:taxonomy}) is stated as a proposition in \S\ref{sec:security-props} and proved by the named result of this appendix; assumptions and preconditions refer to \S\ref{sec:security-ledger}. Parenthesized assumptions are conditional: A8 only for authority-level forward secrecy, A10 only under the in-state peer map (\S\ref{sec:rotation}). CMC is closed by composition with a hardware-rooted runtime.}
\end{table}

\begin{definition}[Epoch-isolation game]\label{def:grfs}
In \(\mathbf{Exp}^{\mathsf{rfs}\text{-}b}(\mathcal{A})\), the adversary outputs equal-shape \((M^0, M^1)\); the challenger runs Phase~I over \(M^b\) with credential \(c^\star\), hands \(\mathcal{A}\) the epoch-0 key material \((\eta_{c^\star}, y_{c^\star}, W_{c^\star}, K)\), performs one honest Phase III.3 full rotation under \(c^\star\) (fresh \(K'\), fresh \(\eta^{\mathsf{next}}_{c^\star}\), \(M' = M^b\)), and hands \(\mathcal{A}\) the post-rotation \(\Sigma'\); \(\mathcal{A}\) outputs a bit. The advantage is defined as in Definition~\ref{def:gcsb}, and the game covers the all-present rewrap policy, as in Appendix~\ref{app:csb}.
\end{definition}

The game captures transit capture of one epoch's wrapping material: \(\mathcal{A}\) holds the full pre-rotation derivation chain of \(c^\star\) and the post-rotation state, and receives no pre-rotation \(\Sigma\), whose compromise is the subject of Theorem~\ref{thm:csb}.

\begin{theorem}[Epoch isolation]\label{thm:rfs}
For every PPT \(\mathcal{A}\), the advantage in Definition~\ref{def:grfs} is at most \(2(\epsilon_{\mathsf{prf}} + \epsilon_{\mathsf{kdf}} + \epsilon_{\mathsf{wrap}}) + \epsilon_{\mathsf{ae}} + 2^{1-\lambda}\).
\end{theorem}
\begin{proof}
The chain of Theorem~\ref{thm:csb} restricted to the post-rotation entries of \(c^\star\), with one adjustment in the PRF step: the reduction must produce \(y_{c^\star} = \mathsf{PRF}_{c^\star}(\eta_{c^\star})\) for \(\mathcal{A}\) while replacing \(\mathsf{PRF}_{c^\star}(\eta^{\mathsf{next}}_{c^\star})\) by a uniform value, which is the standard two-query PRF experiment with one output revealed; it is sound whenever \(\eta^{\mathsf{next}}_{c^\star} \neq \eta_{c^\star}\), an event that fails with probability at most \(2^{-\lambda}\) per direction of the hybrid. The revealed \(W_{c^\star}\) unwraps only the retired entry \([K]_{c^\star}\), which \(\Sigma'\) no longer contains, and the revealed \(K\) is an independent draw that no component of \(\Sigma'\) depends on, so the remaining steps carry over unchanged.
\end{proof}

\begin{proof}[Proof of Proposition~\ref{prop:rfswrap}]
Theorem~\ref{thm:rfs} is the game form of Lemma~\ref{cl:pcfs}: the post-rotation keys \((K', W^\star_{\mathsf{new}})\) are computationally independent of the revealed pre-rotation tuple, since \(K'\) is a fresh draw and the derivation chain of \(W^\star_{\mathsf{new}}\) is rooted in the fresh \(\eta^{\mathsf{next}}_{c^\star}\); the reverse direction, revealing the post-rotation key material against the pre-rotation state, is the same hybrid with the epochs swapped. The claim applies to credentials that have rotated; the dormant-credential and peer-map limitations are catalogued as L3 and L4 (\S\ref{sec:security-limits}), and the plaintext caveat of \S\ref{sec:security-props} is in force when \(M' = M\).
\end{proof}

\begin{proof}[Proof sketch of Proposition~\ref{prop:rfsauth}]
Authority-level forward secrecy is a semantic property of \(E\)'s credential interface and needs no computational game. By (A8), each epoch executes against an authority-bearing secret issued for that epoch and revoked at the next rotation, so authority over a retired epoch requires a secret \(E\) no longer accepts; the same argument runs in reverse for a current-epoch operation under a retired secret. Combined with Theorem~\ref{thm:rfs}, authority is bounded to its issuing epoch.
\end{proof}

\subsection{Coverage and Scope}
\label{app:coverage}

Three boundaries of the model remain informal, each named where it is assumed. The equivalence \(\equiv_E\) and the injectivity of canonical serialization are taken as given (A4, P1); the environment \(E\) and the semantics of \(\mathsf{Release}\) are abstract parameters (P3), so what a conformant result form discloses is a deployment question (L5); and the model covers one custodian and one protected state, with multi-state deployments running independent instances. Mechanized analysis of the games in a tool such as Tamarin or ProVerif is left to future work (\S\ref{sec:security}).

\end{document}